\documentclass[12pt,preprint]{aastex}

\begin{document}
\title{The Large and Small Scale Structures of Dust in the Star-Forming Perseus Molecular Cloud}
\author{Helen Kirk$^{1,2}$, Doug Johnstone$^{2,1}$, and James 
	Di Francesco$^{2,1}$}
\affil{$^1$Department of Physics \& Astronomy, University of Victoria, 
	Victoria, BC, V8P 1A1, Canada; hkirk@uvastro.phys.uvic.ca}
\affil{$^2$National Research Council of Canada, Herzberg Institute of 
	Astrophysics, 5071 West Saanich Road, Victoria, BC, V9E 2E7, 
	Canada; doug.johnstone@nrc-crnc.gc.ca, james.difrancesco@nrc-cnrc.gc.ca}

\begin{abstract}
We present an analysis of $\sim$3.5 square degrees of submillimetre
continuum and extinction data of the Perseus molecular cloud.
We identify 58 clumps in the submillimetre map and
we identify 39 structures (`cores') and 11 associations of
structures (`super cores') in the extinction map.  The cumulative
mass distributions of the submillimetre clumps and extinction cores
have steep slopes ($\alpha \sim 2$ and 1.5 - 2 respectively),
steeper than the Salpeter IMF ($\alpha$ = 1.35), while the
distribution of extinction super cores has a shallow slope
($\alpha \sim$ 1).  Most of the
submillimetre clumps are well fit by stable Bonnor-Ebert spheres with
10~K $<$ T $<$ 19~K and 5.5 $<$ log$_{10}$(P$_{ext}$/k) $<$ 6.0.  The
clumps are found only in the highest column density regions
(A$_{V} >$ 5 - 7 mag), although Bonnor-Ebert models suggest that we
should have been
able to detect them at lower column densities if they exist.
These observations provide a stronger case for an extinction threshold
than that found in analysis of less sensitive observations of the Ophiuchus
molecular cloud \citep{letter}.  The relationship between submillimetre
clumps and their parent extinction core has been analyzed.
The submillimetre clumps tend to lie offset
from the larger extinction peaks, suggesting the clumps formed via
an external triggering event, consistent with previous observations.
\end{abstract}

\section{Introduction}

Molecular clouds require support on their largest scales to prevent
collapse.  The Jeans mass, the maximum mass for which thermal
pressure alone provides sufficient support to counteract gravitational
collapse, is  $\sim$500~M$_{\odot}$ for typical molcular cloud conditions.
Molecular clouds, however, can contain $\gtrsim 10^{4}~$M$_{\odot}$ of
gas, and therefore significant additional support must be
present, unless these clouds are in a state of dynamic collapse.

Several mechanisms have been proposed to account for the difference
between the Jeans mass and total mass in clouds.
In the `standard model' of star formation \citep{Shu, Mestel56, Mous76},
magnetic fields
threading molecular clouds are strong enough to prevent global cloud
collapse, while smaller scale collapse can proceed after
ambipolar diffusion. Recent observations of magnetic field strengths
show that, while important, the fields may not be strong enough
to prevent collapse \citep{Crutcher}.
An alternate mechanism
is that of turbulent support (see MacLow \& Klessen 2004 for a review).
Supersonic motions of large-scale flows are responsible for the prevention
of large-scale collapse, while smaller scale collapse can occur in
regions of flow intersection.  Supersonic line widths have been observed
in molecular clouds \citep[e.g.,][]{Larson}, demonstrating that
turbulent motions are important to consider.
One of the difficulties with the turbulent support model is
the source of the turbulence - without a driving source,
turbulence dissipates quickly due to
shocks, etc.~\citep{ralf}.
The formation of structure within clouds under either of these two support
mechanisms could also be aided by small-scale triggering
\citep[the `globule-squeezing' scenario in][]{Elmegreen},
e.g., through a pressure increase from copious amounts of
ionizing UV radiation from a generation of previously formed O and B stars.
A third option is that the molecular clouds are dynamic
entities without any support.  Large- or intermediate- scale
triggered star formation \citep[e.g., the `collect and collapse' and
`shells and rings' scenarios in][]{Elmegreen} account for the formation
of both the cloud and stars as well as the subsequent rapid dissipation of
the cloud \citep[see, e.g.,][]{Hartmann01}.

Cloud support mechanisms may operate over the entire molecular cloud,
not only sites of ongoing star formation.
Recent developments in submillimetre and IR detectors are allowing for
the column density
structure of molecular clouds to be observed over large (degree)
scales.  Such observations enable, for the first time, the characterization
of a significant fraction of a molecular cloud.
The CO-ordinated Molecular Probe Line Extinction and Thermal Emission
(COMPLETE) Survey \citep[see {\tt http://cfa-www.harvard.edu/COMPLETE};][]{Ridge06}
is one project whose goal is to provide insight into star formation through
large-scale multi-wavelength observations of the Perseus, Ophiuchus,
and Serpens molecular clouds \citep[a subset of the nearby clouds targeted by
the Spitzer c2d Legacy Program; see][]{c2d}.

The Perseus molecular cloud is of particular interest not only because of its
relative proximity, but also because it forms low and intermediate mass
stars, and therefore provides a link between the well-known low mass star
forming Taurus molecular cloud and the massive star forming Orion molecular
cloud.  Regions of studied star formation in Perseus include NGC1333, B1,
IC348, L1448, and B5.

We present observations and analysis of the column density structure
of the Perseus molecular cloud, to provide a basis for
testing the theoretical cloud support mechanisms described above.  We utilize a
combination of (chopped) submillimetre dust continuum observations
to measure the small-scale column density and low resolution stellar
reddening extinction data to measure the large scale column density.
In Section 2, we present our observations and data reduction
techniques, followed by an analysis of the structure (Sections 3 - 7)
including the implications for the models.  We conclude with a summary
in Section 8.


\section{Observations and Data Reduction}\label{sect_obs}

Submillimetre data at 850~$\mu$m of the Perseus molecular cloud were
obtained using the Submillimetre Common User Bolometer Array (SCUBA)
on the James Clerk Maxwell Telescope (JCMT) on Mauna Kea\footnote{The
JCMT is operated by the Joint Astronomy Centre in Hilo, Hawaii on
behalf of the parent organizations Particle Physics and Astronomy
Research Council in the United Kingdon, the National Research Council
of Canada and The Netherlands Organization for Scientific Research.}.
The data we present here are a combination of our own observations
($\sim$1.3 square degrees) with publicly available archival
data\footnote{Guest User, Canadian Astronomy Data Centre, which
is operated by the Dominion Astrophysical Observatory for the National
Research Council of Canada's Herzberg Institute of Astrophysics.} for
a total of $\sim$3.5 square degrees.

Our observations consist of $\sim$400 sq.~arcmin fields mapped using SCUBA's
`fast scan' mode with chop throws of 33\arcsec\ and 44\arcsec\ to complement best
the matrix inversion technique \citep{matrix} for data reconstruction.
The data were taken  in the fall of 2003 on the nights of August 12,
September 3, 18, and 27, and October 1, 4, and 9 under a mean optical depth
at 850~$\mu$m of $\tau = 0.34$, with a variance of 0.01.  The majority the of
archival data was originally presented by \citet{Jenny} and \citet{Sandell01}.

Following the standard procedure, all of the raw data were first flat-fielded
and atmospheric extinction corrected using the normal SCUBA software \citep{Holland}.
To convert the corrected difference measures into an image, we apply the
matrix inversion technique of \citet{matrix}, which was shown to
produce better images from chopped data than other commonly used procedures
such as the Emerson technique for Fourier deconvolution \citep{Emerson79},
often employed at the JCMT.
The matrix inversion technique has several advantages including the
ability to combine data taken with different observing setups (such
as is found in the heterogenous archival data), and weighting measurements
taken under different weather conditions.

The resulting map has a pixel size of 6\arcsec\ and an intrinsic
beamsize of 14\arcsec\ FWHM.
To remove pixel-to-pixel noise which interferes
with the ability to identify clumps properly, the map was
convolved with a FWHM = 14.1\arcsec ($\sigma_{G}$ = 6\arcsec) Gaussian,
producing an effective beam with a FWHM of 19.9\arcsec.

Structures on scales several times larger than the chop throw ($>$120\arcsec)
may be artifacts of the image reconstruction \citep[independent
of the reconstruction technique;][]{matrix}.
We removed these structures through the subtraction of a map convolved
with a Gaussian
with $\sigma_{G}$ = 90\arcsec.  To minimize negative ``bowling'' around bright
sources, all points with values outside of $\pm$ 5 times the mean noise
per pixel were set to $\pm 5$ times the mean noise before convolution.

Since various portions of the map were observed under different weather
conditions (optical depths) and scanning speeds (our observations utilized
fast-scanning, while the archival data were taken with slow-scanning),
the noise across the
final map is not uniform, although it typically varies by only a
factor $< 5$.  The mean and rms
standard deviation are $\sim$10~mJy/bm and $\sim$9~mJy/bm respectively.
Note that the pixels subsample the beam, so that the noise
per pixel is several times larger than the beam.

\begin{figure}[p]
\includegraphics[height=19.5cm, bb=73 -16 538 809]{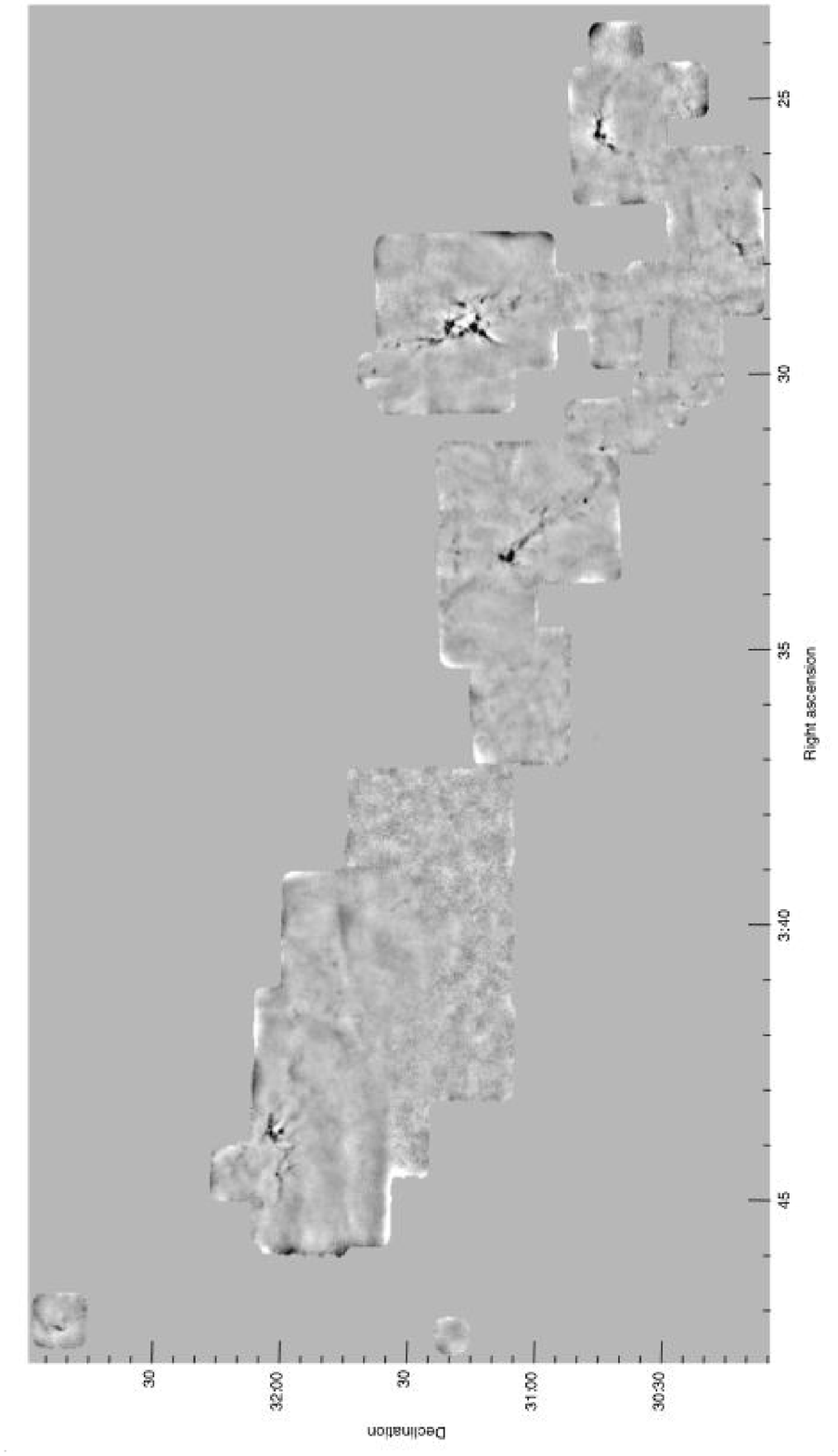}
\includegraphics[height=18.0cm, bb=290 54 322 659]{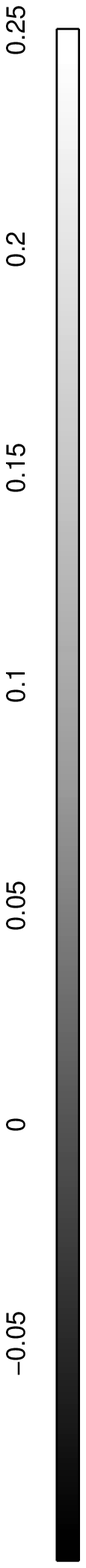}
  \caption{850~$\mu$m observations of the Perseus molecular cloud.
        The scale bar indicates Jy/beam measured at every pixel.
        }
  \label{sub}
\end{figure}

\begin{figure}[p]
\includegraphics[height=18.5cm, bb=79 103 533 690]{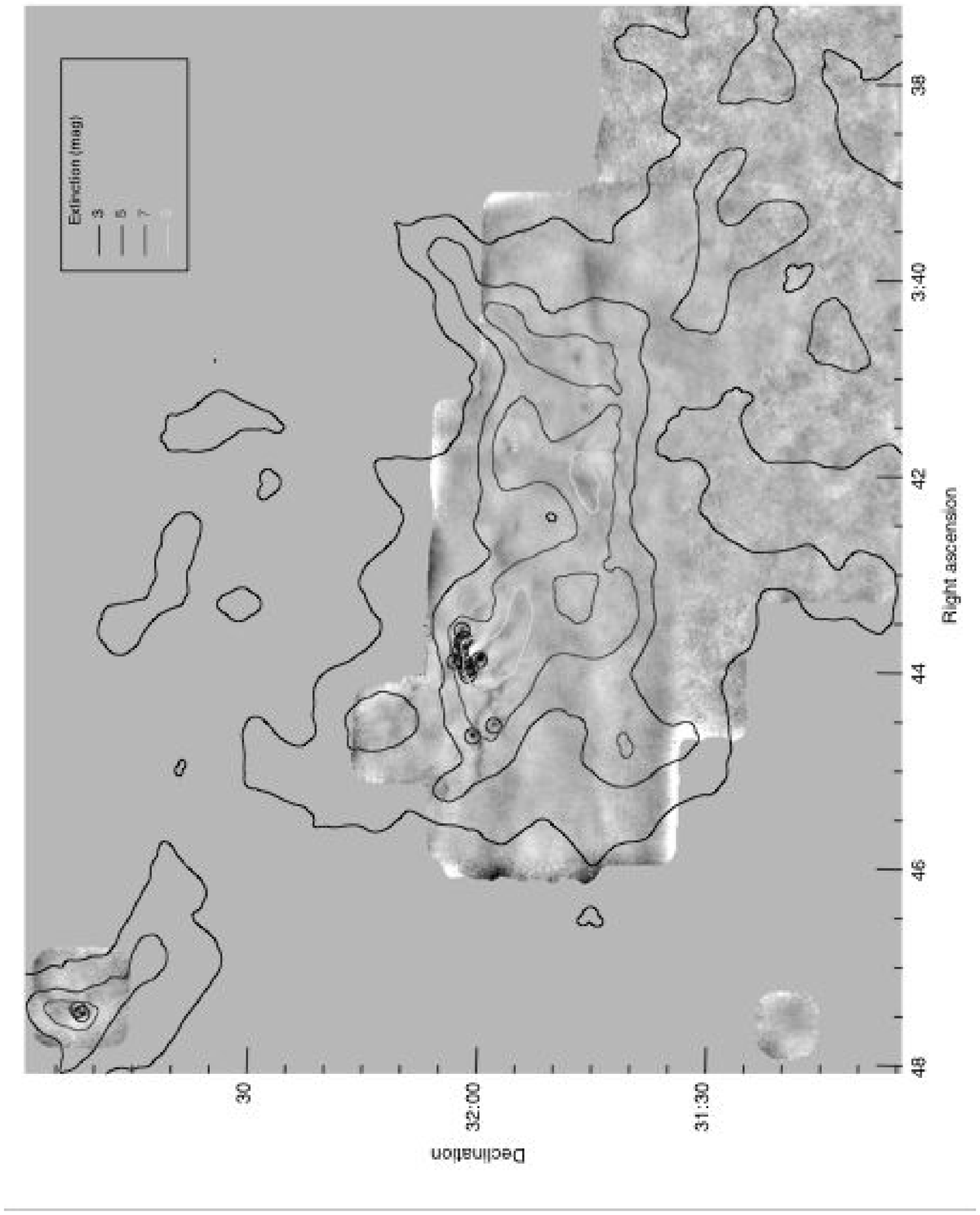}
\includegraphics[height=15.8cm, bb=290 14 322 619]{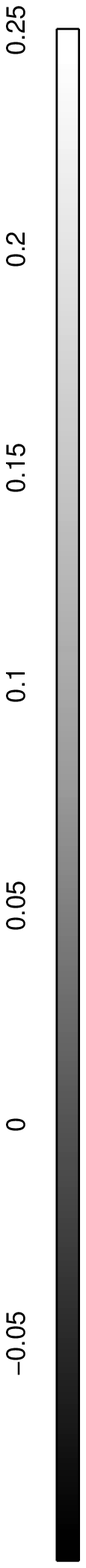}
  \caption{850~$\mu$m observations of the eastern half of Perseus.
        Overlaid are contours denoting the extinction
        \citep[from][see text for further details]{Alves}.
        The dark circles indicate where submillimetre clumps
        were identified (see \S3).  The scale bar indicates
        Jy/beam measured at every pixel.}
  \label{sub+ext1}
\end{figure}

\begin{figure}[p]
\includegraphics[height=18.6cm, bb=99 46 533 747]{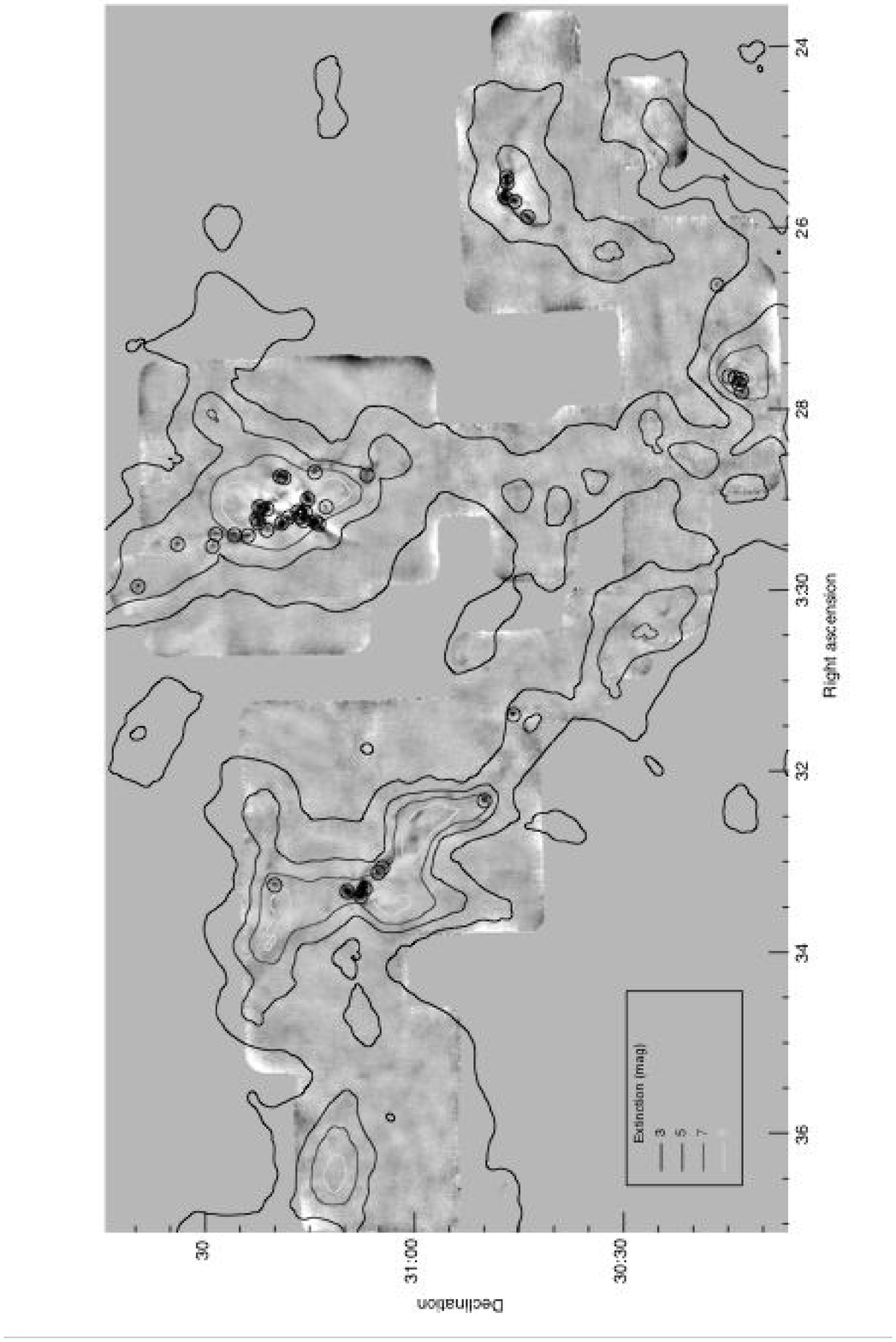}
\includegraphics[height=17.0cm, bb=290 44 322 649]{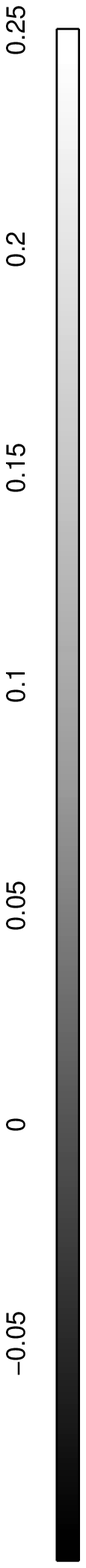}
  \caption{850~$\mu$m observations of the western half of Perseus.
        Overlaid are contours denoting the extinction in magnitudes
        \citep[from][see text for further details]{Alves}.
        The dark circles indicate where submillimetre clumps
        were identified (see \S3).  The scale bar
        indicates Jy/beam measured at each pixel.}
  \label{sub+ext2}
\end{figure}

The resulting map is shown in its entirety in Figure~\ref{sub}, and
Figures~\ref{sub+ext1} and \ref{sub+ext2} show detail of the eastern
and western halves of the surveyed cloud.

The extinction data we present
here were derived from the Two Micron All Sky Survey (2MASS) images of
Perseus by \citet{Alves} \citep[see also][]{Ridge06} using the NICER 
technique \citep{Lombardi01} as a part of the COMPLETE Survey \citep{Goodman04}.
The resolution in the Perseus extinction map is $\sim$2.5\arcmin\ which
has the effect of smoothing out the small scale structure in the map
and diluting regions of high extinction.  Very compact regions of high
extinction could be missed entirely if there is an insufficient number
of background stars detected to contribute significantly to the extinction
calculation.  Although the NICER technique utilizes procedures to remove
embedded and foreground stars from the extinction derivation, this is
difficult, and any still included can introduce errors and affect the
observed cloud morphology.  The extinction data are denoted by
contours overlaid on Figures~\ref{sub+ext1} and \ref{sub+ext2}.
It should be noted that \citet{Schnee05} have shown
that the extinction determined for a region is uncertain at the
0.2~mag level -- a comparison between extinction derived using the NICER
technique with 2MASS data and re-reduced IRAS far-IR data showed this
point-to-point difference even when optimal cloud-specific dust properties
were used. 


\section{Identification of Structure}
We identify structure in both the submillimetre and extinction
data, as described in the following subsections.

\subsection{Submillimetre Continuum}
To identify structure in the 850 $\mu$m map, we used the automated routine
Clumpfind \citep[2D version;][]{clfind,dougoph}.  Some clump-identifying
algorithms assume a pre-determined shape for the structure (typically
Gaussian), leading to the artificial division of more complex objects.
Clumpfind, however, does not assume a shape and instead
utilizes contours to determine clump boundaries.
Bulk clump properties such as the clump mass distribution
have been shown to be similar to those found assuming a Gaussian profile
for clumps \citep{dougoph}.
With Clumpfind, we identify 58 submillimetre clumps down to a level
of 3 times the mean
pixel noise.  Spurious objects that were either smaller
than the beam or noise spikes appearing in regions of higher than average
noise, the vast majority of which occur at the map edges, were
excluded from this total.
Figures~\ref{sub+ext1} and \ref{sub+ext2} illustrate where these
clumps were found.
\citet{Jenny}, hereafter H05, have recently published a similar analysis
on much of the
same region of Perseus using a different method for the data reduction and
the identification of clumps, and our analysis shows good agreement with their
results.  The H05
clumps tend to have slightly higher peak fluxes, as the final submillimetre
map presented in that paper was neither flattened nor smoothed.
Note that a larger number of clumps were identified in H05, as the
clump-detection threshold was lower in their analysis (see 
discussion in $\S4.1$).
Table~\ref{submm_props} lists
the properties of the clumps we identify, along with the corresponding
designations from H05.

The total flux of each clump can be converted into mass assuming that the
emission is optically thin and its only source is thermal emission from
dust.  Following \citet{dougoph},
the conversion is
\begin{equation}
{\rm M_{clump} = 0.23\ S_{850}\left(\exp \Big(\frac{17~K}{T_{d}}\Big) - 1\right)
\ \Big(\frac{\kappa_{850}}{0.02~cm^{2}~g^{-1}} \Big)^{-1}
\ \Big(\frac{D}{250~pc}\Big)^{2}\ M_{\odot}}
\end{equation}
where S$_{850}$ is the total flux at 850~$\mu$m, T$_{{\rm d}}$ is the
dust temperature, $\kappa_{850}$ is the dust opacity at 850~$\mu$m,
and D is the distance.  Following the Spitzer c2d team \citep{c2d},
we adopt a distance of 250 ($\pm$ 50)~pc to the Perseus molecular
cloud as found by \citet{Cernis93} and \citet{Belikov02}.
A range of distances have been estimated for the Perseus molecular
cloud ranging from 350~pc \citep{Herbig83} to 220~pc \citep{Cernis90}, with
several authors suggesting that the system is composed of two distinct
clouds at different distances - e.g., 
\citet{Ungerects87,Goodman90,Cernis03,Ridge05}; 
with the closer cloud being an extension of Taurus  
and the further a shell-like structure.
We also take a typical internal temperature of 15~K and, following
\citet{dougorion}, a dust opacity of $\kappa_{850}$ =~0.02 cm$^{2}$~g$^{-1}$.
Therefore, the conversion factor between Jy and M$_{\odot}$ is 0.48.
H05 adopt values of D = 320~pc, T = 12~K, and
$\kappa$ = 0.012~cm$^{2}$~g$^{-1}$; thus our masses need to be
multiplied by a factor of 4.1 to be compared to the H05 values.
The final masses may be scaled by a factor ranging from $\sim$0.3 to
6 given the uncertainties in distance, dust opacity, and clump
temperature.

In Table~\ref{submm_props}, we estimate the number density for
each clump using the effective clump radii found by Clumpfind,
suggesting that temperatures of tens of Kelvin are
required for only thermal support, assuming ${\rm T_{dust} = T_{gas}}$.

\subsection{Extinction}
\begin{figure}[p]
\includegraphics[height=19.0cm, bb=104 20 508 772]{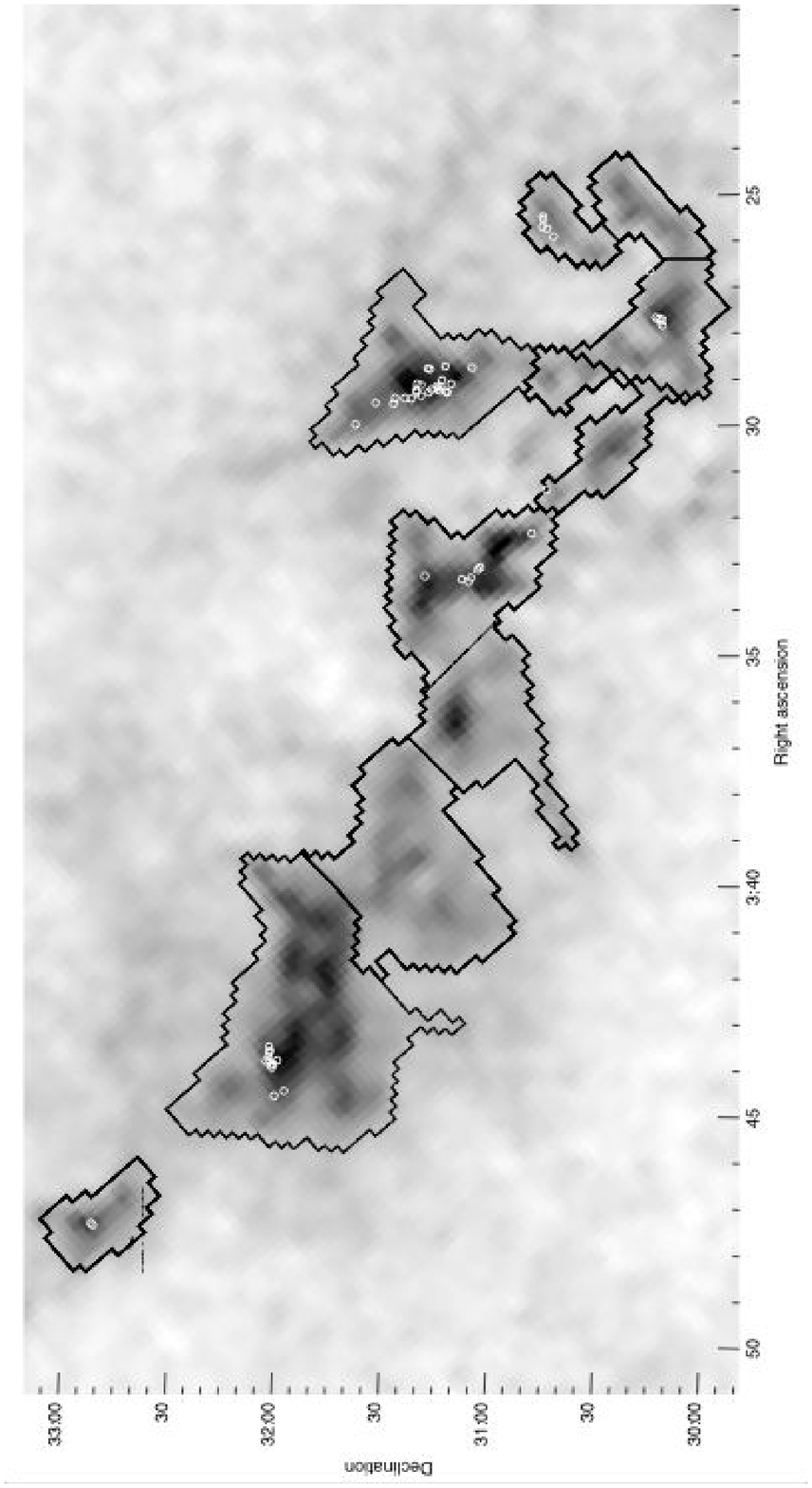}
\includegraphics[height=17.5cm, bb=270 -46 302 738]{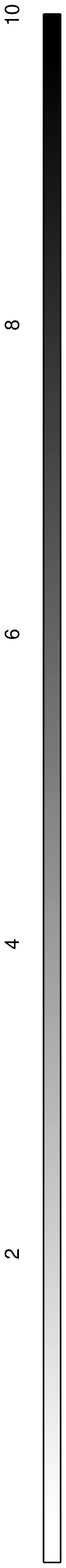}
  \caption{Large scale structure (`super cores') identified in
        extinction data.  The scale bar indicates extinction at each
        point in magnitudes.  The white circles indicate the locations
        of submillimetre clumps identified (see text for details).}
  \label{supcoresdef}
\end{figure}

\begin{figure}[p]
\includegraphics[height=18.6cm, bb=104 20 508 772]{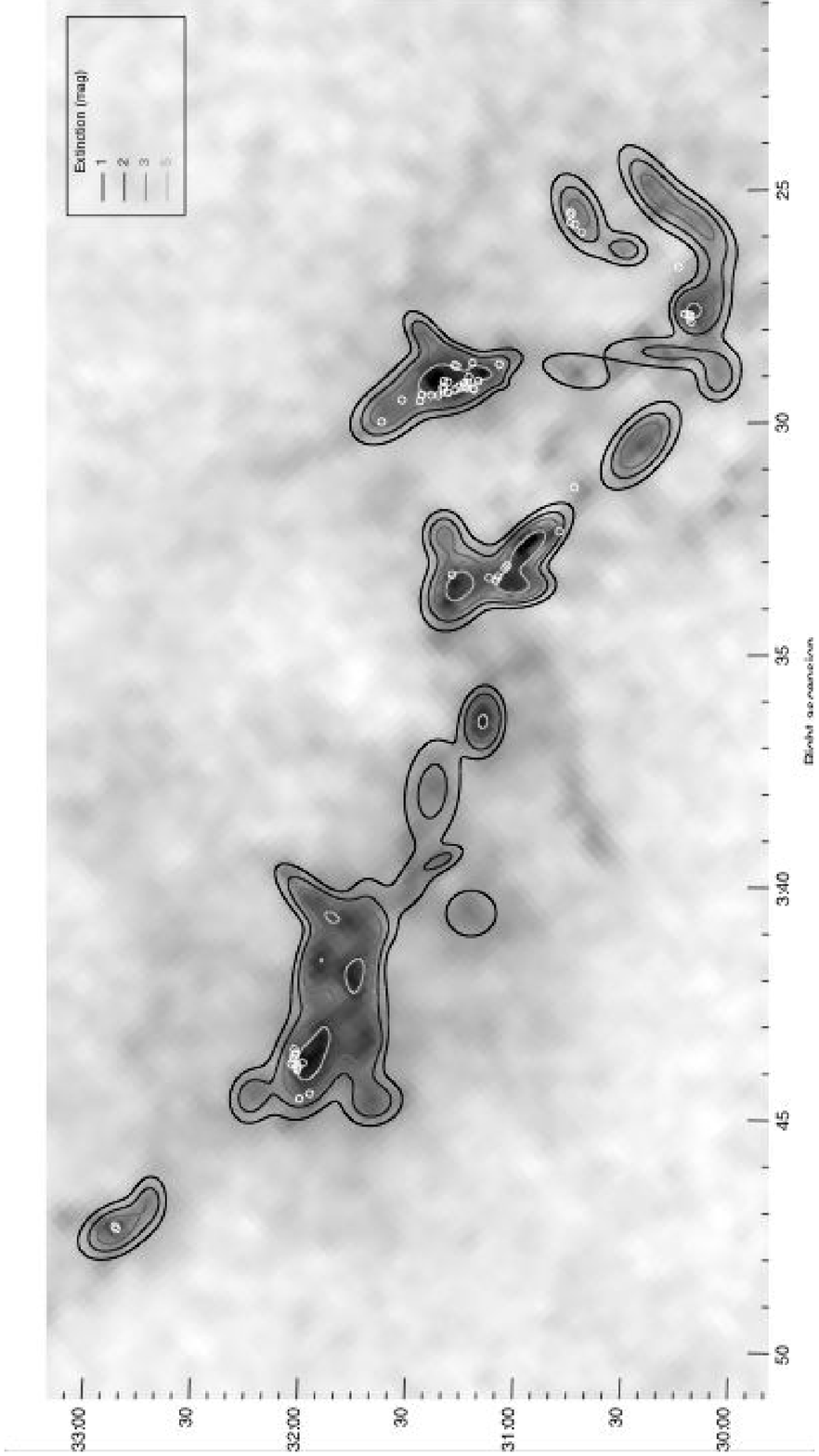}
\includegraphics[height=17.2cm, bb=260 -26 292 758]{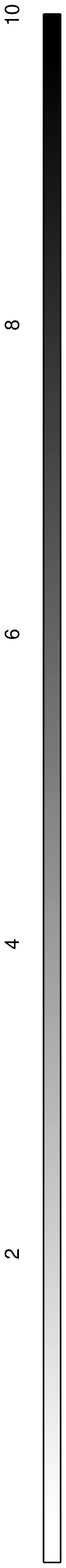}
  \caption{Gaussian model of extinction `cores' identified
        (the contours plotted exclude the background level fit
        to every core).  The scalebar indicates extinction at
        each point in magnitudes.  The white circles indicate the locations
        of submillimetre clumps identified (see text for details).
        }
  \label{coresdef}
\end{figure}

Structure also exists in the extinction maps (see Figs~\ref{sub+ext1} and
\ref{sub+ext2}).  In our submillimetre data, the majority of clumps
were isolated, allowing for easy identification of structure.  In the
extinction map, however, the structure has a large filling factor,
making identification and separation much more difficult.
Visually, structure is apparent on two scales -
a smaller scale consisting of compact objects and a larger scale
consisting of groups of the compact objects within a diffuse background.
Here, we term these two types of structures as cores
and super cores.  The extinction level in the diffuse regions of super cores
is varied so that a simple utilization of the Clumpfind algorithm does
not produce reliable structure identification.  To define the larger
extinction `super cores', we smoothed the data to 5\arcmin\ resolution
and then ran Clumpfind.  The resulting identifications are shown in
Figure~\ref{supcoresdef}.
The smaller extinction `cores' are shaped more regularly, and we
found them to be well fit by 2D Gaussians\footnote{we used the
publicly available IDL mpfit2d routine by Craig Markwardt}.
Although fitting to a Gaussian does have the disadvantage of
assuming a shape for the structure, this procedure allows for a
separation between diffuse background
extinction (associated with the larger extinction super core structure)
and the concentrated extinction in the core region.
Figure ~\ref{coresdef} shows the Gaussian models
of the extinction cores (excluding the background level).

To convert the extinction measures into mass, we adopt a conversion factor
of $<N(H_{I} + H_{2})> / E_{B-V} = 5.8 \times 10^{21}$~atoms~cm$^{-2}$~mag$^{-1}$ \citep{Bohlin78}
and a standard reddening law where A$_{V}$ / E$_{B-V}$ = 3.09 \citep{Rieke85}.
Thus an A$_{V}$ = 1~mag corresponds to a column density of
1.88 $\times  10^{21}$ (H$_{I}$+H$_{2}$) cm$^{-2}$, or
4.40 $\times 10^{-3}$~g~cm$^{-2}$
adopting the standard mean molecular weight $\mu$ = 1.4 \citep{Allens}.
We find the total mass in the extinction map is
${\rm \sim 1.9 \times 10^{4}~M_{\odot}}$
(${\rm \sim 6 \times 10^{3}~M_{\odot} }$ in the region for which we
have submillimetre data).
Previous extinction mass estimates using the Palomar Sky Survey and
mapping in CO have resulted in similar values
- \citet{Bachiller86} estimate 1.2$\times 10^{4}$~M$_{\odot}$
(1.7$\times 10^{4}$~M$_{\odot}$ with their assumed distance of 300~pc),
while \citet{Carpenter00} estimate $\sim$ 0.8 $\times 10^{4}$~M$_{\odot}$
(1.3$\times 10^{4}$~M$_{\odot}$ with their assumed distance of 320~pc) from
\citet{Padoan99}'s observations and H05 estimate
$\gtrsim 0.6 \times 10^{4}$~M$_{\odot}$
($\gtrsim 10^{4}$~M$_{\odot}$ with their assumed distance of 320~pc).

Tables~\ref{extcore_props} and \ref{extsupcore_props} show the
properties of the cores and super cores.  We calculated their number
density using the half width half max from Gaussian fitting and mean radius
derived from the total areal coverage respectively,
and find temperatures of over 50~K are required for purely thermal support,
which is unrealistic.


\section{Mass Distribution}\label{sect_massdistrib}
Previous (sub)millimetre studies of star-forming regions
\citep[e.g.,][]{Motte, dougoph, dougorionb, dougorion, Reid05, Enoch06}
have shown that clumps have a mass distribution well fit by
a broken power law, with the number of clumps with mass above M,
N(M) $\propto$ M$^{-\alpha}$.  The slope $\alpha$ is similar to
or higher than that which characterizes the stellar Initial Mass Function
\citep[$\alpha \sim$ 1.35;][]{Salpeter}, with a turnover to
a shallower power law at low masses.  The turnover observed in
the submillimetre tends to
occur where incompleteness in the clump sample becomes
significant.  The similarity between the clump and initial stellar
mass functions is suggestive of a direct link between the two,
although there are several concerns with this.  The total
submillimetre clump mass often exceeds that expected to be in the final
stars \citep[see][for a review]{Larson05}.  This difference might be
acounted for through inefficient star formation (e.g., if all clumps lost
some percentage of the clump mass reservoir to outflows during formation).
A second problem is that there is no evidence that all clumps
do form stars.  In the turbulent support framework,
a significant number of clumps in fact re-expand \citep[e.g.,][]{turb_evol}.
In most cases, there is insufficient data to determine whether the
clumps are gravitationally bound.
Many theories have been put forth
to account for the rough invariance of the slope including turbulent
fragmentation \citep[e.g.,][]{Padoan04}, competitive accretion
\citep[e.g.,][]{Bonnell05}, and thermal fragmentation
\citep[e.g.,][]{Larson05}.

On larger scales within molecular
clouds, CO observations have shown that structures follow a
mass distribution where the bulk of the mass is contained in the
most massive structures (unlike the submillimetre clumps), with
a lower value for $\alpha$, between 0.6 and 0.8 \citep{Kramer}.
The cause for the difference between the large and small scale mass
distributions is unknown.  It may be due to different methods of
observations, in the definition of structure, the effects of chemistry such as
freeze-out, or to real differences in the structures
at the large and small scales due to the different physical processes
that are responsible for fragmentation.

\subsection{Submillimetre Continuum}
Adopting a common dust temperature of 15~K, we find the submillimetre clumps in
Perseus are well fit by a broken power law with $\alpha \sim 2$, which
falls within the range of slopes found by previous authors, varying
from $\alpha \sim$ 1 - 1.5 \citep{dougoph} to  $\alpha \sim$ 2 \citep{Reid05}.
Using the temperatures calculated from Bonnor-Ebert modelling
(\S\ref{sect_be}), the slope appears to be similar.
A recent study of clumps identified at 1.1~mm in Perseus covering
7.5~sq.\ degrees \citep{Enoch06} identified 122 objects and
found a mass distribution with a similar slope to the one we find, but
includes higher mass objects that belong to more extended structures
than our chopped 850~$\mu$m data are sensitive to.
We find the mass distribution turns over to a power law with 
a shallow slope at
around 0.3 M$_{\odot}$, approximately where our sample begins to suffer from
incompleteness.
Incompleteness becomes important in our sample where clumps have
peak fluxes too low to be identified by Clumpfind.  Clumpfind identified
objects down to a peak of 5 $\sigma$ and extends them to the 3 $\sigma$
level.  Taking a `typical' clump extent of $2.6 \times 10^{-3}$~pc$^{2}$
(1800~arcsec$^{2}$), clumps of masses $\sim$~0.2 to 0.3 M$_{\odot}$
would be missed.
H05 identify clumps to a lower central flux (i.e., a lower cutoff)
and thus find more sources.
Our 1.3 sq.~deg of fast-scan observations
have a higher noise level than the public archival data analyzed in
both papers and we set our Clumpfind threshold higher to reflect this
difference.

\subsection{Extinction}
\begin{figure}[p]
\centerline{\includegraphics[width=16cm]{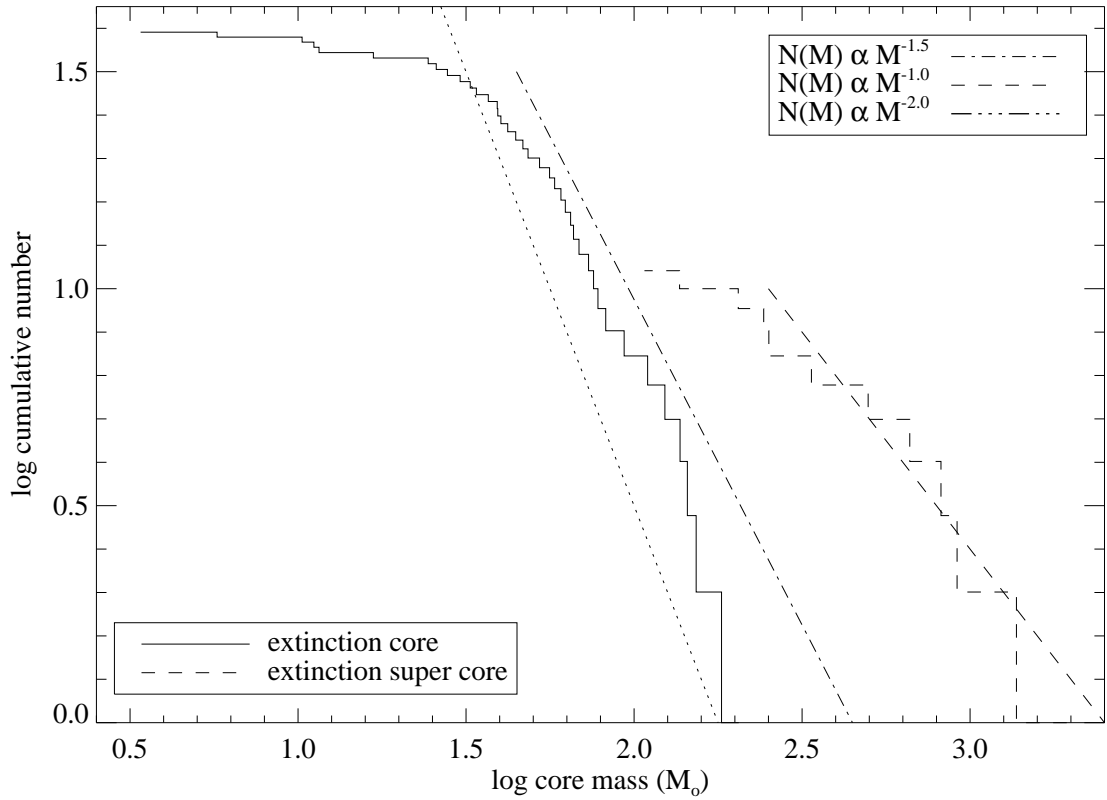}}
  \caption{Cumulative mass distribution for extinction cores and super cores.
        The super cores are well fit by a single power law with a shallow slope,
        while the cores require a steep slope at the high mass end.
	Mass distribution slopes of -1, -1.5, and -2 are shown to 
	guide the eye.}
  \label{massdistrib_ext}
\end{figure}

We also examined the mass distributions of the extinction cores and super
cores (see Figure~\ref{massdistrib_ext}).   The extinction cores have
a similar mass distribution to the submillimetre clumps,
with slope of $1.5 \le \alpha \le 2 $ at the high
mass end and a turnover at $\sim$40 M$_{\odot}$ while the
super cores have a shallower mass
distribution, similar to that seen from CO data, with a slope of
$\alpha \sim 1$.
The small numbers of cores and particularly super cores lead to
greater uncertainties in the derived slopes.
Incompleteness is difficult to quantify accurately for the extinction
data, given the methods used for clump identification.  The extinction
super cores identified comprise virtually all of the mass in the
extinction map at A$_{V} \ge$ 3.  At A$_{V} < 3$, the extinction
is diffuse and unassociated with any apparent extinction structure.  
Thus it is likely that sources of additional
mass have not been missed throughout the majority of the mass range of
identified super cores.  Each of these structures, however, may
be more massive than estimated.  The Gaussian fitting routine for the
extinction cores did not fit all of the extinction above A$_{V}$ = 3, and
thus we expect incompleteness at the lower end of the mass
distribution.  The Gaussian fitting routine models maxima in the
extinction map (i.e., peak plus background level); we stopped our search
at $\sim$5~mag, or a peak of $\sim$3~mag.  Typical core extents are
$\sigma_{G} \sim$~300\arcmin, which corresponds to a mass of 
$\sim$50~M$_{\odot}$.  Thus the turnover we observe at 
$\sim$40 M$_{\odot}$ is probably not real.

The change in behaviour of the extinction core and super core mass
distributions and their close correspondence to the two regimes
previously observed in the submillimetre and CO is intriguing.
It is possible
that the differences are a result of some bias that we introduced
through these definitions of the two sets of objects.
Projection effects may also play a role in the mass distribution
measured.  Most of the cores and super cores are
bounded by similar objects (unlike the submillimetre clumps), making
overlap for the real 3D objects probable.  This may
have a greater effect on the super cores which are less regularly
shaped and larger.  Overlap would lead to larger numbers of massive
objects (i.e., a shallower slope).
With the above caveats, if the slopes are truly different for the
cores and super cores, this may be an indication of the scale over which
fragmentation changes from a top-heavy to a bottom-heavy mass function
due to the importance of different physical processes.


\section{Bonnor-Ebert Modelling}\label{sect_be}
To gain further physical insight into the nature of the clumps, we
model them as
Bonnor-Ebert (BE) spheres -- spherically symmetric structures that are
isothermal, of finite extent, and bounded by an external pressure,
where gravity is balanced by thermal pressure \citep{Bonnor, 
Ebert, Hartmann}.
Previous work \citep[e.g.,][]{Alves01, dougoph, dougorionb, dougorion}
has shown that submillimetre clumps can be well fit by a BE sphere
model.  Each BE sphere is parameterized by its
central density, external pressure, and
temperature, each of which can be extracted from a best fit to the data.

Recent work has shown that caution is needed in the interpretation of the
fit to a BE sphere model, as dynamic entities produced in turbulent
simulations can mimic the column density profile of a stable BE sphere
\citep{BP03}.
BE sphere models are useful, however, in illustrating the minimum
level of support that would be necessary to prevent collapse in an object
of given mass and radius bounded by a finite pressure.

BE spheres are characterized by a one-dimensional radial density profile,
with a family of models defined by each dimensionless truncation radius.
Each family of BE spheres possesses a unique importance
of self-gravity, or equivalently the central concentration of a
clump, with a higher concentration corresponding to a higher importance
of self-gravity.  Each concentration therefore
defines a unique family of BE spheres.
Following \citet{dougorionb}, we define the concentration to be
(in terms of observable quantities)
\begin{equation}
{\rm C = 1 - \frac{1.13\,B^{2}\,S_{850}}{(\pi R^{2}_{obs})\,f_{0}}} 
\end{equation}
where B is the beamsize, S$_{850}$ is the total flux,
$\rm R_{obs}$ the radius, and f$_{0}$ the peak
flux.   The concentrations are approximate due to the relatively large
size of the beam compared with the clump radius as well as projection
effects.   To be fit by the Bonnor-Ebert
sphere model, the concentration
must be greater than 0.33 (corresponding to a uniform density sphere) and
less than 0.72
\citep{dougoph}.
To be stable from collapse, clumps with concentrations above 0.72
require additional support mechanisms (for example, pressure from
magnetic fields).  Alternatively, these high concentration clumps
may be more evolved and already experiencing collapse.
\citet{Josh} and \citet{Josh06} analyzed multiwavelength data in the
B1 core of Perseus and found that all clumps with concentrations above
0.75 contained protostars, while none of those at low concentrations
($<$ 0.4) and few at the intermediate concentrations did.
Concentration thus appears to
be a good indicator of time evolution.

We use the concentration to fit the clumps to stable Bonnor-Ebert spheres
following \citet{dougorion}, again adopting a distance of 250~pc and an
opacity of 0.02~cm$^{2}$~g$^{-1}$.  Non-thermal support may exist in the
clumps.  \citet{Goodman98} find in their survey of prestellar cores that
non-thermal support levels are a non-negligible fraction of the thermal
support level.  Following
\citet{dougorion}, we assume an equal level of thermal and non-thermal
support.  Comparable levels of thermal and non-thermal support may
not be the case everywhere.  \citet{Tafalla04} show
that non-thermal support is negligible in the cores in the quiescent
star-forming region Taurus.  A lower level of non-thermal support
would increase our best-fit temperatures.

Our best fits include temperatures ranging from 10~K to 19~K and external
pressures within 5.5 $\le$ log$_{10}$(P/k) $\le$ 6.0 (see
Table~\ref{submm_props}).
The physical properties fit to clumps with concentrations outside
the stable range are less reliable.
The external clump pressures found with the BE modelling are in the
range expected to be generated by the pressure due to the weight of the
surrounding material in the molecular cloud.  For example, the pressure
due to overlying material can be written as
${\rm P/k = 1.7 \times 10^{4}~\overline A_{V}A_{V}~cm^{-3}~K}$,
where the mean extinction is $\sim$2.2~mag (see \S\ref{sect_ext_thresh})
and thus the mean pressure in the molecular cloud
is log$_{10}$(P/k) $\approx$ 4.9.  The highest extinction in the
cloud is 11.8~mag, corresponding to a central pressure in
the cloud due to the surrounding material of log$_{10}$(P/k)
$\approx$ 5.6.  Higher pressures can be generated by the extra
weight of material in large cores with local mean extinctions
higher than 2.2~mag.
The pressures fit here are lower than
those found by \citet{dougoph} in the Ophiuchus molecular
cloud (they found 6~$\le$~log$_{10}$(P/k)~$\le$~7),
consistent with the fact that the total column
densities (or extinctions), and hence pressures found in Ophiuchus are
larger -- \citet{letter} (hereafter JDK04) found a mean extinction of
4~mag and a peak extinction of 35~mag.

The temperatures derived from the Bonnor-Ebert fits provide
a second estimate of the clump masses (also included in
Table~1), as discussed in Section \ref{sect_massdistrib}.
These clump masses correspond closely to those calculated assuming
a constant temperature of 15~K, indicating that the temperature assumed
should not be of critical importance to the further analysis presented.

\section{Clump Environment: Extinction Threshold}\label{sect_ext_thresh}

\begin{figure}[p]
\centerline{\includegraphics[width=16cm]{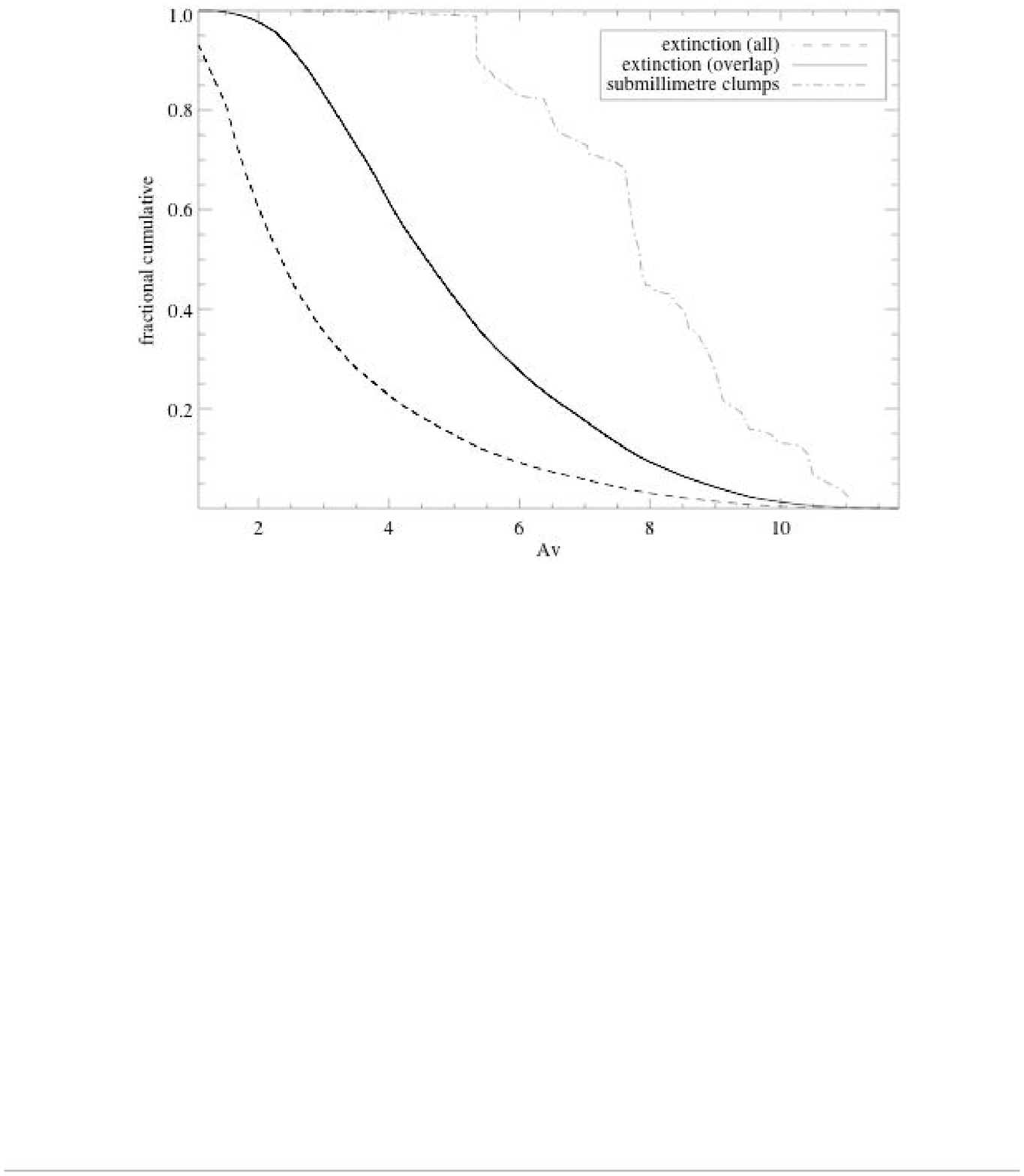}}
  \caption{Cumulative mass in extinction and submillimetre map versus 
	extinction.  The dashed line indicates the extinction over the 
	entire region of Fig.~\ref{sub}, while the solid line indicates
        the extinction only in the region of Fig.~\ref{sub} where
        our submillimetre data exist.}
  \label{cumulative_ext}
\end{figure}

Following JDK04, we examine the relationship between
the locations of small-scale, submillimetre clumps relative to the overall
cloud structure traced by extinction.
Figure~\ref{cumulative_ext} shows the cumulative mass in the submillimetre
clumps and extinction map at increasing A$_{V}$.  Table~\ref{ext_bin}
includes the fraction of the submillimetre clumps and extinction data
within three bins of extinction.  These demonstrate that most of the
mass of the cloud lies at low extinction -- 58$\%$ at A$_{V}\le$ 5 --
the submillimetre clump mass is biased towards the high extinction regions
and only a small fraction (1$\%$) is found at A$_{V}\le$ 5.
Very little of the mass of the cloud is at high
extinction -- 1$\%$ at A$_{V} \ge$ 10, while 13$\%$ of the submillimetre
clump mass is found there.
The portion of Perseus for which we have submillimetre observations is in
itself biased towards higher extinctions - the extinction data of the entire
region of Figure~\ref{sub} suggest 86$\%$ of the mass of the cloud
is at A$_{V} <$ 5 and a mere 0.4$\%$ at A$_{V} >$ 10.

These data suggest small scale structure and hence stars
are able to form only at higher extinctions and here we argue
that this is not an observational bias.  We model the effect of
lower extinction (and therefore lower external pressure)
on the observability of clumps using the BE sphere model, following the
procedure of JDK04.

We use the extinction as a measure of the local pressure and determine
how a model clump's observable properties would be expected to vary with
extinction for a given importance of self-gravity (or equivalently
concentration, C) assuming it to be a BE sphere.
Following \citet{McKee}, the pressure at depth {\it r} is
P({\it r}) = ${\rm \pi G \overline \Sigma \Sigma}(r)$ where $\overline \Sigma$ is
the mean column density and $ \Sigma(r)$ the column density measured
from the cloud surface to depth {\it r}.  Near the centre of the cloud,
$\Sigma(r) \simeq \Sigma(s)/2$, where $\Sigma(s)$ is the column
density through the cloud at impact parameter {\it s}.
The extinction can be expressed as ${\rm A_{V} = (\Sigma / \Sigma_{0})}$ mag
where ${\rm \Sigma_{0} = 4.68 \times 10^{-3}~g~cm^{-2}}$ \citep{McKee}.
The pressure in the cloud at depth {\it r} can then be approximated as
P($r$)/k = 1.7${\rm \times 10^{4} \overline A_{V} A_{V}}(s)~{\rm cm^{-3}~K}$, where
${\rm \overline A_{V}}$ is the mean extinction through the cloud
(which we take to be 2.2~mag) and
${\rm A_{V}}(s)$ is the extinction through the cloud at impact parameter {\it s}.
For a given concentration, the mass and radius
of a BE sphere scale as ${\rm M_{BE}\ \propto\ P^{-1/2}}$ and
${\rm R_{BE}\ \propto\ P^{-1/2}}$ \citep{Hartmann}, and thus the
column density scales as ${\rm \Sigma_{BE}\ \propto\ P^{1/2}}$.
If we assume a constant temperature and dust grain opacity, then for
a given concentration, observable clump properties scale as follows --
${\rm S_{850}\ \propto\ A_{V}}(s)^{-1/2}$, ${\rm f_{0}\ \propto\ A_{V}}(s)^{1/2}$,
and R ${\rm \propto\ A_{V}}(s)^{-1/2}$, where S$_{850}$ is the total flux,
f$_{0}$ is the peak flux, and R the radius.  All three quantities increase
with increasing concentration.

\begin{figure}[p]
\centerline{\includegraphics[width=16cm]{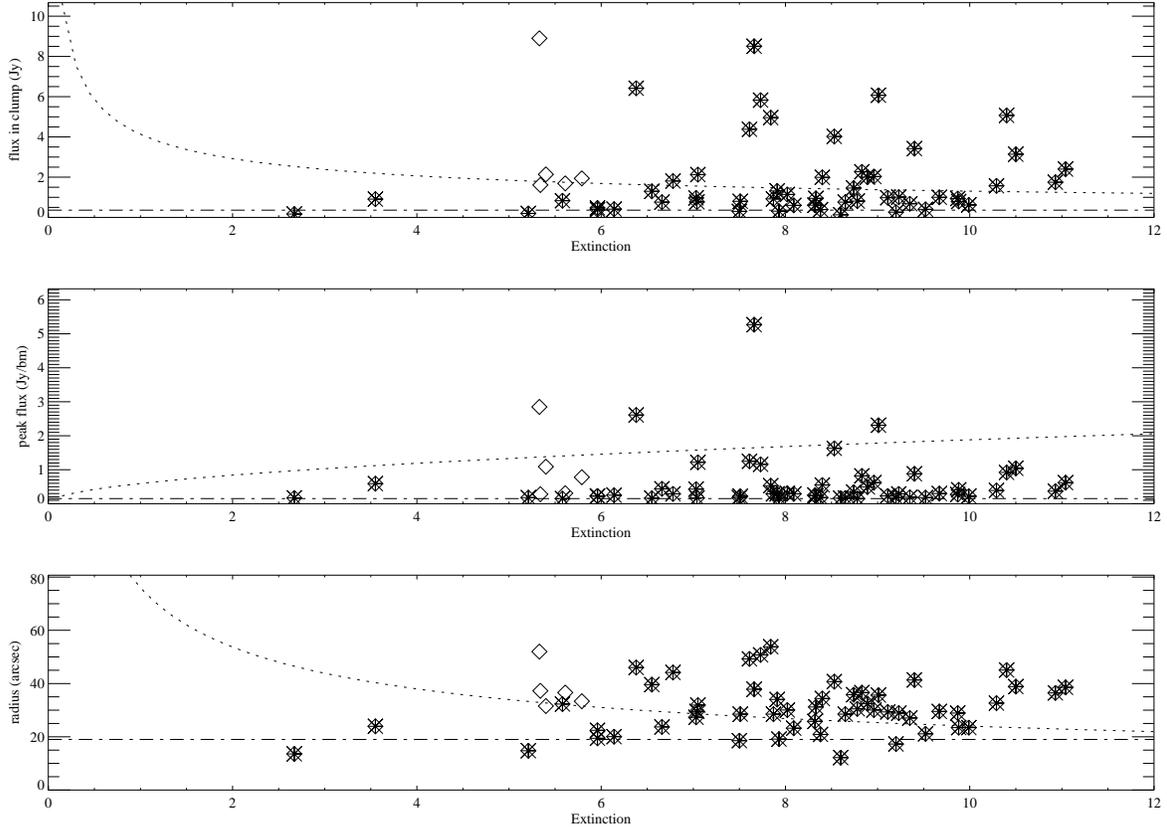}}
  \caption
        {Extinction threshold of submillimetre clumps.  Each clump's total
        flux, peak flux, and radius are plotted versus the extinction
        at that location.  The dotted line indicates the relationship
        expected for a Bonnor-Ebert sphere model with each of the clump
        properties.  The dash-dot line
        denotes the observational thresholds below which our identification
        of clumps becomes incomplete.  The observational threshold for the total
        flux is derived similarly to the incompleteness level for the
        mass distribution, as discussed in \S4.1.  Diamonds
        represent all of the clumps while the asterisks denote those
        not found in the L1448 region (see text for details).
}
  \label{extthresh}
\end{figure}

Figure~\ref{extthresh} plots these quantities versus the local extinction.
The dotted line illustrates the BE model relations for a concentration
matching the observed clumps at A$_{V}$ = 5.  The lack of clumps at lower
$A_{V}$ is incompatible with our detection of clumps at higher $A_{V}$,
suggesting an extinction threshold in clump formation.
If all of the submillimetre clumps detected are considered, the extinction
threshold appears to be at A$_{V} \sim$ 5.  If we ignore the
clumps in L1448 (non-asterisked diamonds), however, we find an extinction
threshold at A$_{V} \sim$ 7.
L1448 (extinction core \#38) is unusual in that it is within the only
extinction core with a low value of peak extinction that contains
submillimetre clumps.  Previous studies have shown that L1448 contains
several very powerful outflows which contain at least
as much momentum as that found in the quiescent cores in the region
(estimated from the mass and small turbulent velocity of the cores)
and more energy than the gravitational binding energy of the region
\citep{WC00}.  If there is a mechanism in place for transferring
some of the momentum and energy into the surrounding core
material, much of that material would dissipate,
leading to the lower peak exinction observed in the region.
To include consideration of this possibility, we denote its associated
clumps with a
different symbol in the extinction threshold analysis to allow considerations
both with and without them.  In the rest of our analyses, we find the
submillimetre clumps in L1448 exhibit similar properties to those
in other cores.
A similar argument for advanced evolution might also be put forth for the
NGC1333 region, well known for its active star formation and
numerous protostellar outflows.  Here,
we merely note that the extinction properties of submillimetre clumps
within the L1448 core appear
different from the other extinction cores and offer the above evolution
argument as a possible reason for this difference.

The archival submillimetre data included in the present analysis were
also examined by H05 who compared them with
C$^{18}$O observations to measure the large scale structure of the
cloud and search for a column
density threshold.  H05 examined the distribution of submillimetre
clumps (identified through a contouring procedure rather than Clumpfind)
with respect to the background column density inferred from C$^{18}$O
in terms of a
probability distribution weighted towards how frequently each column
density occurs in the map.  They found
a sharply decreasing likelihood of submillimetre clumps
at lower extinctions, but a non-zero probability for their lowest extinction
bin.
The sharp decrease in probability is in agreement with the results
we present in Table~\ref{ext_bin}, although H05 find ten
clumps at A$_{V} <$ 5.2, contrary to our results (we identify two, which
may be noise).
Generally, our clumps do correspond well to those
found by \citeauthor{Jenny}, with the differences mainly due to definitions
of clump boundaries and identification thresholds which are not easily
comparable given the different data reduction methods.
Our two clumps at low extinction corresponds to two of
their ten.  A further two of the ten do not correspond to
any submillimetre clumps in our map.
These two unmatched clumps are located in regions where
there is no submillimetre structure in our map above the threshold we set for
identification and visually there appears to be only
noise features.  The isolation of these two clumps from regions containing
the bulk of the clumps coupled with the lack of visible
submillimetre structure in our map further suggests that these are
noise features.  The remaining six clumps are located in regions
where we see clumps, but the extinction we measure
at these locations is much higher than found by H05 (two clump locations
are at 5.5 - 6 mag, while the other four are around 8-9 mag in our map).
The majority of the disagreement in clump identification at low
column densities is due to the different methods of calculating column
density and extinction.  Our method (using extinction from 2MASS) 
is less prone to
chemical effects such as freeze-out at high densities or photodisociation 
at the cloud edge which could effect H05's C$^{18}$O map and 
arguably produces a more accurate count of the total column density
of material along the line of sight, although our data does have
lower resolution (2.5\arcmin\ vs 1\arcmin).  Furthermore, our extinction map
includes any foreground dust along the line of sight and could be distorted by
young embedded protostars (see discussion in \S\ref{sect_obs}).
We note that \citet{Enoch06} found an extinction threshold of
$\sim$5~mag in their analysis of 1.1~mm observations of the Perseus
molecular cloud.

Column density or extinction thresholds for star formation have been found
in other regions.  For example, our previous work (JDK04) used the same method
of analysis as outlined above to argue that an extinction threshold of
A$_{V} \sim 15$ exists in the Ophiuchus molecular cloud.  In addition,
\citet{Onishi} analyzed C$^{18}$O observations of the Taurus
molecular cloud in combination
with IRAS data and argued that recently formed protostars (``cold IRAS
sources'') are found only in regions with a column density of
N(H$_{2}$) $> 8 \times 10^{21}~{\rm cm}^{-2}$, corresponding to an
A$_{V}$ of $>$ 4 (adopting the conversion factor used in this paper; \S3.2),
and that {\it all} regions with column densities above this threshold
contain `cold IRAS sources', suggesting that once the critical column density
is reached, formation occurs quickly.  In contrast, we find several
extinction regions (cores \#11, 20, and 25 or super core \#4) in
Perseus above our extinction threshold of A$_{V}$ = 5 that do not contain any
submillimetre clumps.  This may be a reflection of the different environment
in which stars form in the two clouds, or it may an observational effect
(i.e.\ sensitivity or the definition of cores).
No objects from the IRAS point source catalog lie within the empty
super core \#4, although counterparts to the lowest flux objects in
Taurus could be missed since Taurus is roughly three-fifths
as distant as Perseus
\citep[140 $\pm$ 10 pc;][]{Kenyon94}.  We also do not see any
hint of submillimetre clumps in the empty super core, even below
our detection threshold, implying that any clumps present would not be
comparable to those observed in the other cores in Perseus.
The definition of extinction cores is another possibility for
an observational difference between the Taurus and Perseus surveys.
The angular resolutions are comparable (the \citeauthor{Onishi} maps have a
beamsize of 2.7\arcmin, while our extinction maps have a resolution of 2.5\arcmin)
so the Taurus extinction cores could be further
subdivided on a physical scale, however, chemical processes such as freeze-out
could make the CO distribution smoother and hence less clumpy.
Empty super core \#4 (which contains core \#20), however, appears to be
rather isolated and it
is unlikely that a smoothing effect would result in a substantial
change in its definition.

An extinction threshold is a natural consequence of the magnetic support
model.  Under this model, dense clumps become gravitationally
unstable through ambipolar diffusion.  The ambipolar diffusion timescale
is a function of the density of ions in the region and only becomes
short enough to be significant for higher extinctions where cosmic rays
are the sole source of ionization.  In particular, \citet{McKee} argues
for an extinction threshold between 4 and 8~mag, the exact value depending on
the density of the cloud and the characteristic density in which
cosmic rays dominate the ionization process.  Cloud geometery and
the strength of the local interstellar radiation field
also play a role in the column density threshold observed.
The extinction thresholds observed in Perseus and Taurus are easily
consistent with that predicted by the magnetic support model.  Given
the effect of geometry, etc., the Ophiuchus results also appear to
be consistent, as argued by JDK04.
Further research is required to determine if the turbulent support model
is able to provide an explanation for an extinction threshold.


\section{Triggered Star Formation}
\begin{figure}[p]
\centerline{\includegraphics[width=16cm]{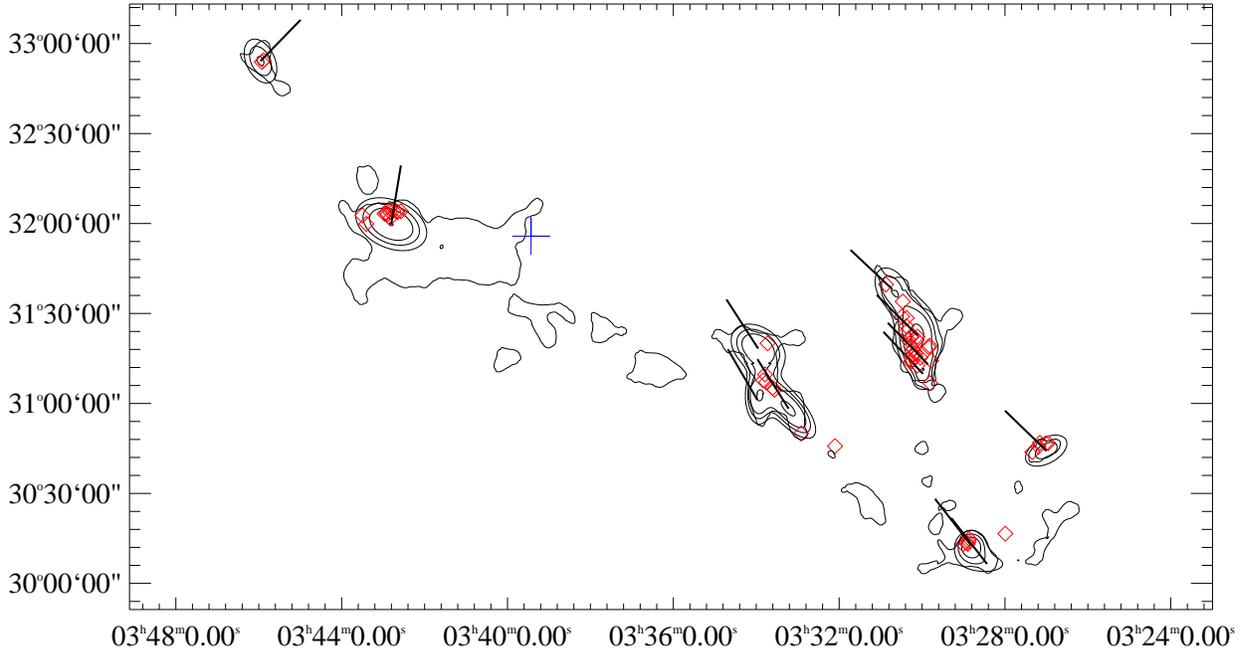}}
  \caption{Possible evidence for triggering in the Perseus molecular cloud.  The
        background contour indicates the A$_{V}$ = 5 level in the cloud to aid
        in orientation.  Other contours represent Gaussian fits to the
        extinction cores containing significant numbers of submillimetre
        clumps (diamonds).  Vectors denote the direction of the B star
        40 Persei, which
        has been suggested as a trigger for star formation in the region (see
        text).  The plus indicates the centre of the
        `Perseus ring' \citep{Ridge05}, an unrelated feature.}
  \label{triggering}
\end{figure}

\begin{figure}[p]
\centerline{\includegraphics[width=16cm]{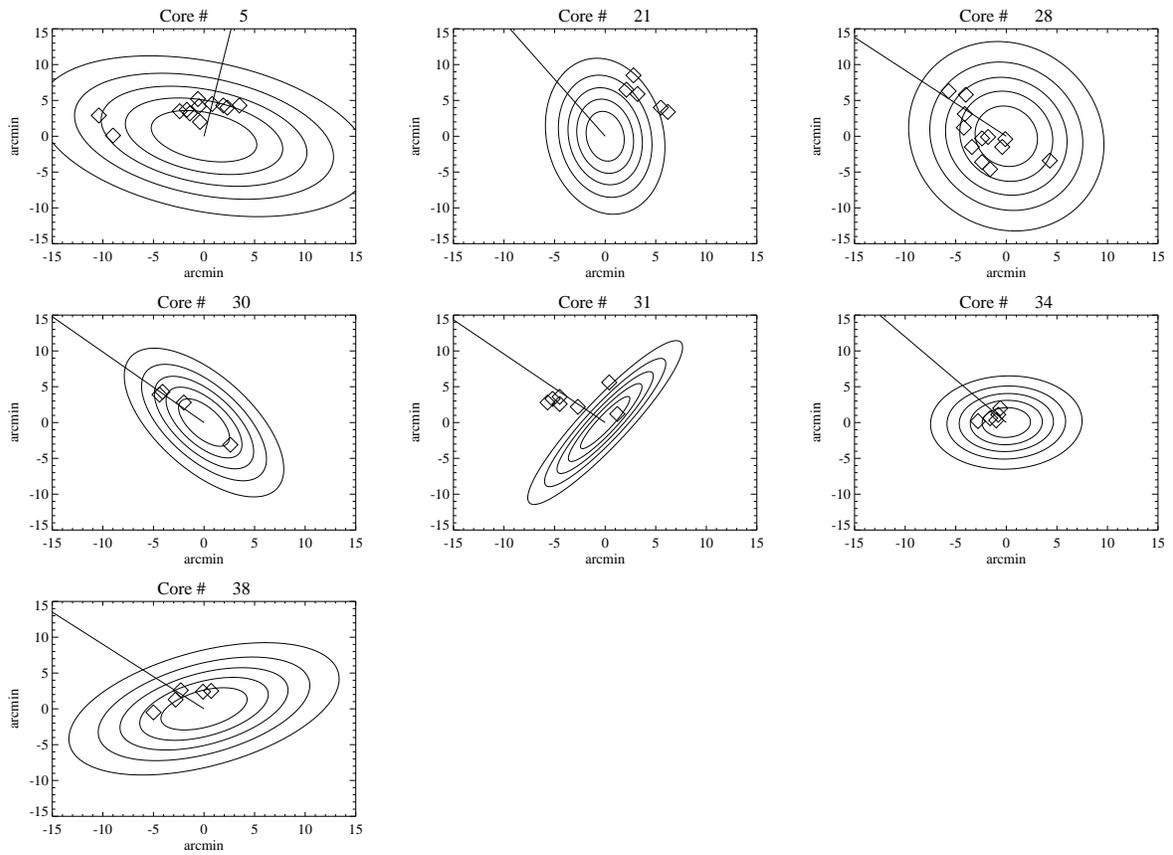}}
  \caption{The distribution of submillimetre clumps within extinction
        cores.  Contours show the Gaussian model fit to each
        extinction core containing more than two submillimetre clumps.
        The vectors indicate the direction to 40 Per.}
  \label{trigger_indiv}
\end{figure}

Figure~\ref{triggering} illustrates
how the submillimetre clumps are preferentially located offset from the peak
of their parent extinction core (see also Figure~\ref{trigger_indiv}
for a close-up of the distribution of submillimetre clumps within
each extinction core).
The difference in resolution between the two data sets is substantial
($\sim$20\arcsec\ vs. 2.5\arcmin), but this is not the cause of the
offset in peak positions.  For example, we smoothed the submillimetre data to
a resolution of 2.5\arcmin\ and found that the offset still remained.
The offsets across extinction
structure furthermore appear to be correlated, as discussed in more
detail below.
One would expect, from simple models of
either magnetic or turbulent support, for clumps to
preferentially form in the densest regions.
While our observations are only able to demonstrate that the submillimetre
structure is offset from the peak {\it column} density, it is
difficult to imagine a scenario where geometrical effects alone
produce the correlation in submillimetre clump offsets while having no
relation to offsets in the underlying density distribution.
Thus, the submillimetre offsets suggest an additional mechanism to
magnetic or turbulent support may be at work.
The correlation between the submillimetre clump offsets from
the peaks in each
extinction core over our entire map suggests the importance of
an outside agent in clump formation or evolution.

A triggered formation scenario for the Perseus molecular cloud has
been previously suggested by \citet{Walawend04}.  There, infrared observations
of a cometary cloud in the L1451 region (to the southwest of our map),
that appears to be eroding by UV radiation, led to the suggestion
that the B0.5 star 40 Persei, of the Perseus OB
assocation, is a likely candidate for triggering star formation in that region.
The small-scale (`globule-squeezing') formation scenario is also consisent
with our observations of Perseus as a whole.  For example, UV radiation from
members of the OB association could erode and heat the surfaces of
extinction cores on the side facing the radiation, leading to
a triggering of the evolution of pre-existing structure on that
side of the core.  

A moderate temperature gradient across the
extinction structures caused by incoming UV radiation is likely
insufficient to explain the visibility of submillimetre clumps
solely on one side of the extinction cores.  Submillimetre clumps
are detected with fluxes an order of magnitude above the 3$\sigma$ noise
level.  Temperatures differing by factors of more than four
would be required to explain the non-detections versus detections
on opposing sides of the core.  This is unlikely to be the case as
our BE models suggest that the clumps we detect have temperatures of
$\sim$15~K, and thus any equivalent population of undetected submillimetre
clumps would be at temperatures of $\sim$4~K.

\begin{figure}[p]
\centerline{\includegraphics[width=16cm]{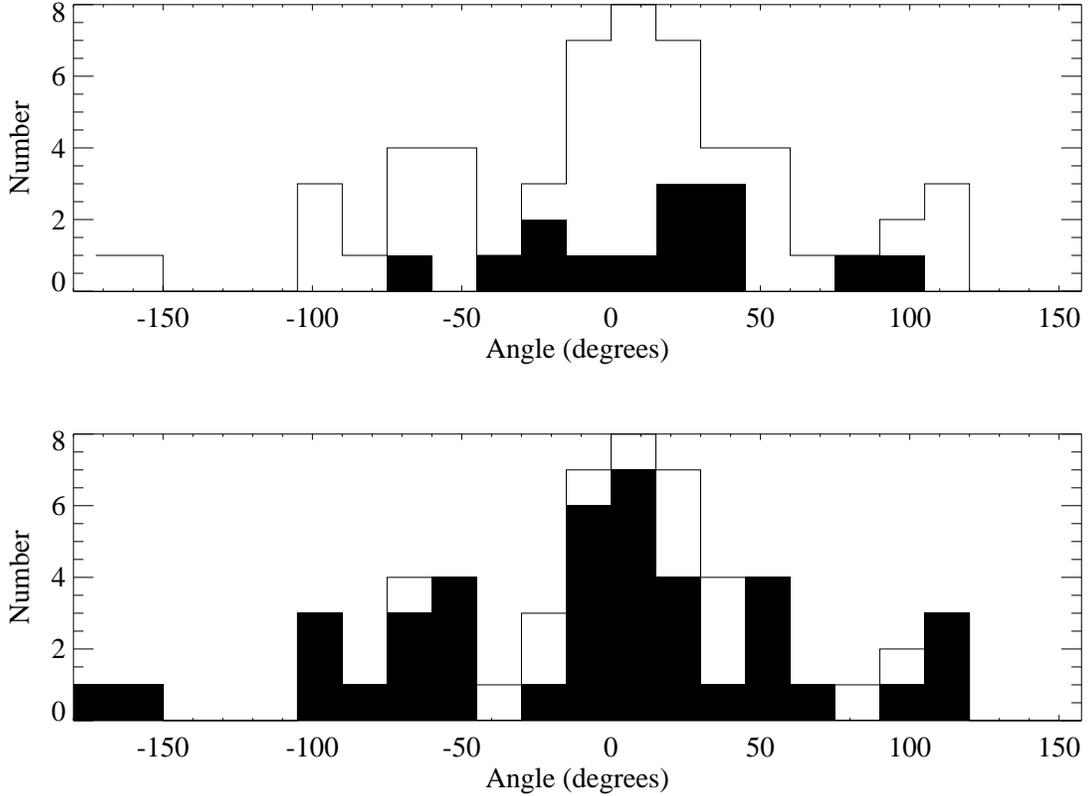}}
  \caption{The distribution of angles between the
        submillimetre clumps and the separation vector between the
        extinction core centre and 40 Per,
        showing that the clump angles are in broad agreement.
	The top plot shows the distribution for all clumps (solid line)
	and those eastward of 40 Per (shaded), while the bottom shows
	those westward of 40 Per.
        The distribution of clump angles for those eastward and
        westward of 40 Per do not appear significantly different,
        confirming this location is consistent with being the trigger.}
  \label{trigger_angle}
\end{figure}

The vectors in Figure~\ref{triggering} indicate the direction
of 40 Per from each of the relevant extinction cores, illustrating
that while the correspondence between the clump locations and
the direction to 40 Per is not perfect, the two are in good agreement
especially given that we are only viewing a 2D projection the region.
We can quantify the agreement between submillimetre clump locations
and the scenario of 40 Per as a trigger as follows.  Assuming (for
simplicity) that all the
extinction cores were spherical, we would expect that triggering
would take place where incident radiation from 40 Per hits the core,
i.e., within $\pm$ 90\arcdeg\ of the separation vector between the
extinction core and 40 Per.  Here we will refer to the angle that
clumps are offset from the separation vector between their parent extinction
core centre and 40 Per as the clump angle.  Examining our submillimetre clumps,
we find that 78$\%$ have clump angles within the $\pm$ 90\arcdeg\ boundary and
86$\%$ within $\pm$ 100\arcdeg, with an average clump angle of -11\arcdeg\ (see
Figure~\ref{trigger_angle} for the distribution).
Several clumps have angular separations well outside the $\pm$ 90\arcdeg\
range - all of these are located in the western portion of the map where
the extinction core geometry is more complex.  These clumps may be more
poorly described as associated with their Gaussian model extinction core,
or otherwise require three dimensional geometry to fully understand them.
The distribution of clump angles is
generally in agreement with 40 Per as a candidate trigger, but
does not necessarily rule out other potential trigger candidates.
A trigger source nearby 40 Per would be expected to
give a similar distribution of clump angles over a range of
extinction cores, but with clumps in cores westward of 40 Per having
preferentially negative clump angles and clumps eastward of 40 Per
having preferentially positive clump angles, or vice versa.  We examined the
distribution of clump angles for those eastward and westward of
40 Per and do not observe such a skew (see Figure~\ref{trigger_angle}).
Clumps eastward of 40 Per
have an average clump angle of 13\arcdeg\ to 40 Per and clumps westward
of 40 Per have an average clump angle of -19\arcdeg\ to 40 Per, but the
overall distributions do not appear to be skewed, and the difference in the
averages may in part be due to small number statistics.  Thus 40 Per appears
to be the likely trigger, though any nearby (similar angle from the cloud)
sources are not ruled out.
Other possibilities for this geometric coincidence exist such as
clump motion from the parent extinction
core caused by stellar wind pressure perhaps from a source such as 40 Per.

We note that the ring observed in $^{13}$CO measurements of the Perseus molecular
cloud as a part of the COMPLETE survey \citep{Ridge05} appears to
be unrelated to any triggering event due to its location
(Figure~\ref{triggering}).  This is consistent with the \citet{Ridge05}
results that the ring appears to be behind the bulk of the cloud.

One outstanding puzzle is the extinction cores which show no evidence
of clumps, especially the ones with significant portions above the
extinction threshold (\#11, 20, 25) as discussed earlier
(Section~\ref{sect_ext_thresh}).
The three dimensional geometry of the extinction cores is not
known, but this could explain the lack of submillimetre clumps seen in
these `empty' extinction cores.  Shielding by a large column of
cloud material separating the empty extinction cores from the
trigger is possible.
Higher signal to noise observations of the `empty' extinciton cores may
reveal submillimetre clumps, but any such clumps would possess a lower mass
and central density than the clumps presented here.  This would still
lead to the question of why two extinction cores with similar
peak extinctions would develop different populations of submillimetre clumps.
Higher sensitivity observations of the `empty' extinction cores (especially
extinction core \#20 which does not lie within a larger super core in which
clumps are observed) are required to understand what processes make them
different.  Spitzer c2d observations of these regions should provide
some clue to their protostellar content.


\section{Conclusions}

We present an analysis of 3.5 square degrees of submillimetre continuum
data and the corresponding extinction map of the Perseus molecular cloud.
We identify structure in both maps of submillimetre emission (clumps) and
extinction (cores and super cores).  The cumulative mass distribution of all
three sets of structures are well characterized by broken power laws.
The submillimetre clumps and extinction cores have high-end mass
distribution slopes of  $\alpha \sim 2$ and 1.5 - 2 
respectively.  These
are slighly steeper than the Salpeter IMF ($\alpha \sim$ 1.35), but within
the range found in submillimetre clumps in other star forming regions.
In contrast, the mass distribution of extinction super cores is best fit by a
shallow slope of $\alpha \sim 1$, corresponding to slopes observed in structures
identified in large scale CO maps.  The difference between the extinction
cores and super
cores may be an observational bias or it may indicate the scale over
which different processes become important in fragmentation.

The majority of submilimetre clumps can be well fit by a Bonnor-Ebert
sphere model of equal thermal and non-thermal internal pressure,
with external pressures ranging from
${\rm 5.5 <\ \log_{10}(P_{ext}/k) < 6.0}$ and temperatures ranging
from 10 to 19~K.  The derived pressures are comparable to the
pressures expected to be exerted by the weight of the surrounding
cloud material and the temperatures fall
within the range expected for a molecular cloud.

We show that small scale (submillimetre) structure (clumps) is located only in regions
of high extinction.  Submillimetre clumps are found only at A$_{V}~>$ 5 - 7,
although BE models suggest that we should have been able to detect clumps
at lower A$_{V}$ had they existed.  In turn, this suggests that clumps are
only able to form above a certain extinction level.  An extinction threshold
is consistent with the model of magnetic cloud support, where the timescale
for ambipolar diffusion is only of a reasonable length in regions above
an A$_{V}$ of 4 - 8 mag.  It is less clear if the turbulent support model
can explain our observations.

The submillimetre clumps were preferentially found offset from the peaks
of several extinction cores.  The correlation of these locations
suggests a small-scale triggering event formed the submillimetre
clumps in the region.  Furthermore, the position of the young B0 star
40~Per, previously suggested as a source of triggering for the region
by \citet{Walawend04},
coincides with the expected position of a triggering source.

Large-scale submillimetre surveys, such as the one presented here, will
become practical to carry out on a large number of molecular clouds with
SCBUA-2, a bolometer array to be installed at the JCMT in late 2006.
SCUBA-2 will have a higher sensitivity and a larger field of view than SCUBA,
allowing large areas to be mapped up to 1000 times faster than using SCUBA.
Legacy surveys have been approved, including the Gould's Belt Survey, in
which nearby molecular clouds will be mapped in their entirety (A$_{V} >$ 1)
in submillimetre continuum, as well as more focussed complimentary observations
of molecular line emission and polarimetry.
These will enable a comprehensive study of the large scale environment of
molecular clouds, allowing a determination of the importance of the cloud
support mechanisms.


\section{Acknowledgements}
We thank Joao Alves and Marko Lombardi for providing the extinction data as 
well as
fruitful discussions on its interpretation.  We would also like
to thank Jenny Hatchell for allowing us to see the data presented
in her paper before publication.  We are also grateful for the helpful
comments provided by the referee, Alyssa Goodman.

HK is supported by a Univeristy of Victoria Fellowship and a 
National Research Council of Canada GSSSP Award.  DJ is supported 
by a Natural Sciences and Engineering Research Council of Canada 
grant.

\markboth{{\it Bibliography}}{{\it Bibliography}}


\clearpage
\begin{deluxetable}{cccccccccccccc}
\rotate
\tabletypesize{\scriptsize}
\tablewidth{0pt}
\tablecolumns{14}
\tablecaption{Properties of submillimetre clumps in Perseus.\label{submm_props}}
\tablehead{
\colhead{Name \tablenotemark{a}} &
\colhead{RA \tablenotemark{b}} &
\colhead{Dec \tablenotemark{b}} &
\colhead{f$_{0}$ \tablenotemark{c}} &
\colhead{S$_{850}$ \tablenotemark{c}} &
\colhead{R$_{eff}$ \tablenotemark{c}} &
\colhead{Mass\tablenotemark{d}} &
\colhead{Conc\tablenotemark{e}} &
\colhead{Temp\tablenotemark{e}} &
\colhead{M$_{BE}$\tablenotemark{e}}&
\colhead{log $n_{cent}$\tablenotemark{e}} &
\colhead{log P$_{ext}$/k\tablenotemark{e}} &
\colhead{H05 \tablenotemark{f}} &
\colhead{Extinction \tablenotemark{g}}
\\
\colhead{(SMM J)} &
\colhead{(J2000.0)} &
\colhead{(J2000.0)} &
\colhead{(Jy/bm)} &
\colhead{(Jy)} &
\colhead{(\arcsec)} & 
\colhead{M$_{\odot}$} &
\colhead{} &
\colhead{(K)} &
\colhead{M$_{\odot}$} & 
\colhead{cm$^{-3}$} &
\colhead{cm$^{3}$~K$^{-1}$} &
\colhead{\#} &
\colhead{Core \#}
 }
\startdata
034769+32517&   3:47:41.6&  32:51:48.0&  0.29&  0.60&  23.&   0.3& 0.45&  12.& 0.54&  5.2&  5.9& 78&  1\\
034764+32523&   3:47:38.8&  32:52:18.9&  0.25&  0.61&  26.&   0.3& 0.48&  10.& 0.66&  5.2&  5.8& 79&  1\\
034472+32015&   3:44:43.7&  32:01:32.3&  0.44&  0.76&  24.&   0.4& 0.56&  10.& 0.84&  5.5&  5.9& 14&  5\\
034461+31587&   3:44:36.8&  31:58:46.1&  0.18&  0.30&  19.&   0.2& 0.31&  14.& 0.19&  4.9&  5.9& 19&  5\\
034410+32022&   3:44:06.1&  32:02:17.7&  0.19&  0.68&  27.&   0.4& 0.30&  17.& 0.32&  4.6&  5.7& 22&  5\\
034404+32025&   3:44:02.8&  32:02:30.5&  0.23&  0.98&  29.&   0.6& 0.29&  18.& 0.39&  4.6&  5.8& 18&  5\\
034402+32020&   3:44:01.3&  32:02:00.8&  0.27&  0.94&  29.&   0.5& 0.41&  14.& 0.56&  4.8&  5.8& 16&  5\\
034396+32040&   3:43:57.7&  32:04:01.6&  0.22&  0.80&  28.&   0.4& 0.36&  17.& 0.36&  4.6&  5.7& 17&  5\\
034395+32030&   3:43:57.2&  32:03:01.8&  0.82&  2.27&  37.&   1.3& 0.71&  12.& 1.92&  5.6&  5.6& 13&  5\\
034394+32008&   3:43:56.5&  32:00:50.0&  1.04&  3.14&  39.&   1.7& 0.72&  13.& 2.24&  5.7&  5.6& 12&  5\\
034385+32033&   3:43:51.0&  32:03:21.2&  0.34&  1.46&  36.&   0.8& 0.52&  12.& 1.26&  5.1&  5.7& 15&  5\\
034376+32031&   3:43:45.8&  32:03:10.4&  0.15&  0.81&  30.&   0.4& 0.17&  17.& 0.37&  4.5&  5.7& --&  5\\
034373+32028&   3:43:43.9&  32:02:52.9&  0.17&  0.76&  28.&   0.4& 0.22&  17.& 0.35&  4.6&  5.7& 26&  5\\
034363+32032&   3:43:38.3&  32:03:12.1&  0.17&  0.31&  19.&   0.2& 0.29&  14.& 0.19&  4.8&  5.9& 23&  5\\
033335+31075&   3:33:21.3&  31:07:34.6&  1.16&  5.83&  51.&   3.2& 0.72&  15.& 3.33&  5.5&  5.5&  2& 21\\
033329+31095&   3:33:17.9&  31:09:34.3&  1.25&  4.38&  49.&   2.4& 0.79&  13.& 2.95&  5.5&  5.5&  1& 21\\
033326+31069&   3:33:16.1&  31:06:58.2&  0.53&  4.96&  54.&   2.8& 0.54&  15.& 2.66&  4.9&  5.6&  4& 21\\
033322+31199&   3:33:13.4&  31:19:58.0&  0.18&  0.39&  21.&   0.2& 0.29&  15.& 0.22&  4.8&  5.9& 82& 22\\
033309+31050&   3:33:05.4&  31:05:03.4&  0.17&  0.77&  30.&   0.4& 0.26&  17.& 0.36&  4.5&  5.7&  6& 21\\
033303+31044&   3:33:02.2&  31:04:27.1&  0.16&  1.30&  40.&   0.7& 0.26&  18.& 0.52&  4.3&  5.5&  5& 21\\
033229+30497&   3:32:18.0&  30:49:47.1&  1.22&  2.13&  32.&   1.2& 0.76&  12.& 1.73&  5.8&  5.8& 76& 23\\
033134+30454&   3:31:20.9&  30:45:28.4&  0.59&  0.91&  24.&   0.5& 0.62&  10.& 0.98&  5.7&  5.9& 77& 23\\
032986+31391&   3:29:51.8&  31:39:08.0&  0.25&  0.41&  20.&   0.2& 0.42&  12.& 0.36&  5.1&  6.0& --& 26\\
032942+31283&   3:29:25.3&  31:28:21.2&  0.18&  0.20&  15.&   0.1& 0.28&  13.& 0.14&  5.0&  6.1& 64& 28\\
032939+31333&   3:29:23.5&  31:33:20.8&  0.23&  0.39&  19.&   0.2& 0.36&  15.& 0.21&  4.8&  5.9& 58& 26\\
032931+31232&   3:29:18.6&  31:23:14.0&  0.29&  0.92&  28.&   0.5& 0.44&  12.& 0.70&  5.0&  5.8& 63& 28\\
032930+31251&   3:29:18.0&  31:25:07.8&  0.29&  1.81&  44.&   1.0& 0.55&  11.& 1.64&  5.0&  5.5& 57& 28\\
032928+31278&   3:29:17.4&  31:27:49.7&  0.22&  0.47&  22.&   0.3& 0.40&  13.& 0.35&  4.9&  5.9& 61& 28\\
032925+31205&   3:29:15.1&  31:20:31.3&  0.18&  0.40&  21.&   0.2& 0.29&  15.& 0.23&  4.8&  5.9& 70& 28\\
032919+31131&   3:29:11.4&  31:13:06.7&  2.61&  6.42&  46.&   3.6& 0.83&  16.& 3.27&  5.6&  5.7& 42& 30\\
032917+31184&   3:29:10.6&  31:18:24.5&  0.88&  3.42&  41.&   1.9& 0.68&  13.& 2.34&  5.5&  5.6& 46& 28\\
032917+31217&   3:29:10.3&  31:21:42.4&  0.39&  1.57&  33.&   0.9& 0.46&  14.& 1.00&  5.0&  5.8& 54& 28\\
032916+31135&   3:29:10.0&  31:13:30.4&  5.27&  8.51&  38.&   4.7& 0.84&  19.& 3.22&  5.9&  6.0& 41& 30\\
032914+31152&   3:29:08.9&  31:15:12.2&  0.55&  1.99&  34.&   1.1& 0.56&  13.& 1.49&  5.3&  5.8& 51& 31\\
032912+31218&   3:29:07.5&  31:21:53.8&  0.37&  1.75&  36.&   1.0& 0.50&  13.& 1.25&  5.0&  5.7& 56& 28\\
032911+31173&   3:29:06.9&  31:17:23.8&  0.29&  1.01&  29.&   0.6& 0.41&  15.& 0.58&  4.8&  5.8& 62& 28\\
032910+31156&   3:29:06.6&  31:15:41.7&  0.62&  2.03&  30.&   1.1& 0.49&  15.& 1.14&  5.2&  6.0& 50& 31\\
032905+31149&   3:29:03.3&  31:14:59.1&  0.49&  2.00&  32.&   1.1& 0.45&  16.& 1.03&  5.0&  5.9& 52& 31\\
032905+31159&   3:29:03.2&  31:15:59.0&  2.31&  6.07&  36.&   3.4& 0.70&  17.& 2.72&  5.8&  6.0& 43& 31\\
032901+31204&   3:29:01.0&  31:20:28.5&  0.92&  5.07&  45.&   2.8& 0.61&  16.& 2.71&  5.3&  5.7& 45& 28\\
032900+31119&   3:29:00.3&  31:11:58.5&  0.16&  0.13&  12.&   0.1& 0.24&  12.& 0.11&  5.2&  6.1& 65& 30\\
032899+31215&   3:28:59.6&  31:21:34.2&  0.62&  2.40&  39.&   1.3& 0.63&  12.& 1.89&  5.4&  5.7& 47& 28\\
032891+31145&   3:28:54.9&  31:14:33.4&  1.63&  4.02&  41.&   2.2& 0.79&  14.& 2.53&  5.7&  5.7& 44& 31\\
032866+31179&   3:28:40.2&  31:17:54.3&  0.31&  1.14&  30.&   0.6& 0.42&  14.& 0.69&  4.9&  5.8& 55& 31\\
032865+31060&   3:28:39.2&  31:06:00.3&  0.15&  0.83&  32.&   0.5& 0.25&  17.& 0.39&  4.4&  5.6& 75& 30\\
032865+31185&   3:28:39.2&  31:18:30.0&  0.29&  1.32&  34.&   0.7& 0.44&  14.& 0.87&  4.8&  5.7& 60& 28\\
032861+31134&   3:28:36.8&  31:13:29.6&  0.43&  0.97&  27.&   0.5& 0.57&  11.& 1.01&  5.4&  5.8& 49& 31\\
032780+30121&   3:27:48.4&  30:12:08.7&  0.23&  0.95&  31.&   0.5& 0.39&  15.& 0.53&  4.7&  5.7& 37& 34\\
032771+30125&   3:27:42.8&  30:12:31.4&  0.30&  1.00&  30.&   0.6& 0.45&  12.& 0.77&  5.0&  5.8& 36& 34\\
032766+30122&   3:27:40.0&  30:12:12.7&  0.22&  0.62&  23.&   0.3& 0.27&  17.& 0.29&  4.7&  5.9& 40& 34\\
032765+30130&   3:27:39.0&  30:13:00.4&  0.41&  0.79&  23.&   0.4& 0.50&  11.& 0.73&  5.3&  6.0& 35& 34\\
032763+30139&   3:27:38.0&  30:13:54.2&  0.18&  0.26&  17.&   0.2& 0.30&  14.& 0.17&  4.9&  6.0& 39& 34\\
032662+30153&   3:26:37.3&  30:15:20.8&  0.17&  0.17&  14.&   0.1& 0.23&  13.& 0.12&  5.1&  6.1& 80& 35\\
032581+30423&   3:25:49.1&  30:42:18.1&  0.28&  1.61&  37.&   0.9& 0.41&  16.& 0.84&  4.7&  5.7& 32& 38\\
032564+30440&   3:25:38.7&  30:44:03.0&  1.09&  2.14&  32.&   1.2& 0.72&  12.& 1.72&  5.8&  5.8& 29& 38\\
032560+30453&   3:25:36.2&  30:45:20.2&  2.85&  8.90&  52.&   4.9& 0.84&  17.& 4.01&  5.6&  5.6& 28& 38\\
032543+30450&   3:25:26.0&  30:45:05.2&  0.31&  1.69&  37.&   0.9& 0.42&  16.& 0.90&  4.7&  5.7& 31& 38\\
032537+30451&   3:25:22.3&  30:45:10.0&  0.78&  1.94&  33.&   1.1& 0.68&  12.& 1.68&  5.7&  5.7& 30& 38\\
\enddata

\tablenotetext{a}{Name formed from J2000 positions (hhmm.mmdddmm.m)}
\tablenotetext{b}{Position of peak flux within clump (accurate to 6\arcsec).}
\tablenotetext{c}{Peak flux, total flux, and radius derived from {\it clfind} \citep{clfind}.  Note a beamsize of 19.9\arcsec is used for the peak flux.}
\tablenotetext{d}{Mass derived from the total flux assuming T$_{d}$ = 15~K and
        $\kappa_{850} = 0.02\,$cm$^{2}$~g$^{-1}$, d = 250~pc.}
\tablenotetext{e}{Concentration, temperature, mass, central number 
        density, and external pressure derived from Bonnor-Ebert 
        modelling (see text).}
\tablenotetext{f}{Best corresponding submillimetre clump in \citet{Jenny}.  
	More clumps were identified in their survey, as discussed in
	$\S$4.1.}  
\tablenotetext{g}{Closest corresponding extinction core.}

\end{deluxetable}

\clearpage
\begin{deluxetable}{cccccccccccccc}
\tabletypesize{\small}
\tablewidth{0pt}
\tablecolumns{10}
\tablecaption{Properties of extinction cores in Perseus.\label{extcore_props}}
\tablehead{
\colhead{Ref \#} &
\colhead{RA \tablenotemark{a}} &
\colhead{Dec \tablenotemark{a}} &
\colhead{Peak \tablenotemark{b}} &
\colhead{A$_{0}$ \tablenotemark{b}} &
\colhead{Mass \tablenotemark{b}} &
\colhead{$\sigma_{x}$ \tablenotemark{b}} &
\colhead{$\sigma_{y}$ \tablenotemark{b}} &
\colhead{$<$n$>$ \tablenotemark{b}} &
\colhead{Extinction \tablenotemark{c}} 
\\
\colhead{} &
\colhead{(J2000.0)} &
\colhead{(J2000.0)} &
\colhead{(A$_{V}$)} &
\colhead{(A$_{V}$)} &
\colhead{(M$_{\odot}$)} &
\colhead{(\arcsec)} &
\colhead{(\arcsec)} &
\colhead{($10^{3}$cm$^{-3}$)} &
\colhead{Sup.~Core \#}
}
\startdata
  1&   3:47:43.8&  32:52:08.0&  4.1&  2.6&  80.3& 394.& 253.&  5.2&  6\\
  2&   3:47:01.8&  32:42:34.9&  2.3&  2.4&  24.8& 192.& 291.&  3.9&  6\\
  3&   3:44:47.1&  31:40:31.6&  2.9&  3.4&  47.8& 327.& 258.&  4.4&  2\\
  4&   3:44:42.4&  32:15:09.2&  2.7&  2.8&  39.2& 311.& 237.&  4.4&  2\\
  5&   3:43:54.2&  31:58:53.4&  6.0&  3.7& 204.9& 540.& 324.&  5.5&  2\\
  6&   3:43:38.3&  31:43:51.3&  3.2&  4.0&  60.0& 431.& 227.&  3.5&  2\\
  7&   3:43:25.5&  31:41:24.2&  1.1&  3.7&   3.4& 144.& 115.&  3.6&  2\\
  8&   3:43:08.7&  31:54:33.6&  1.6&  3.5&  11.7& 147.& 252.&  3.1&  2\\
  9&   3:42:57.8&  31:48:16.5&  0.9&  3.3&  10.9& 435.& 138.&  0.8&  2\\
 10&   3:42:01.4&  31:48:04.8&  4.1&  4.3& 143.5& 561.& 321.&  3.5&  2\\
 11&   3:41:48.3&  31:57:43.0&  3.6&  3.0&  66.3& 482.& 198.&  3.1&  2\\
 12&   3:41:34.7&  31:43:21.4&  2.3&  2.8&  33.2& 398.& 184.&  2.6&  2\\
 13&   3:40:45.2&  31:48:47.0&  2.5&  4.6&  43.3& 189.& 469.&  2.2&  2\\
 14&   3:40:37.3&  31:14:12.6&  1.9&  3.4&  45.6& 333.& 364.&  2.5&  7\\
 15&   3:40:26.6&  31:43:13.5&  1.9&  3.2&  34.7& 255.& 362.&  2.7&  2\\
 16&   3:40:17.5&  31:59:50.6&  2.8&  2.8&  56.0& 201.& 502.&  2.4&  2\\
 17&   3:40:01.1&  31:31:10.8&  1.6&  3.8&  26.8& 213.& 403.&  1.9&  7\\
 18&   3:39:26.7&  31:21:44.6&  1.9&  4.2&  11.4& 118.& 264.&  3.2&  7\\
 19&   3:37:57.6&  31:25:20.6&  2.6&  2.8& 103.3& 629.& 327.&  1.9&  7\\
 20&   3:36:26.1&  31:11:12.6&  5.3&  3.6&  70.9& 208.& 332.&  7.9&  4\\
 21&   3:33:31.0&  31:01:11.3&  5.3&  2.2&  65.8& 346.& 185.&  7.3&  3\\
 22&   3:33:29.2&  31:18:14.1&  5.9&  2.7& 144.7& 430.& 292.&  6.9&  3\\
 23&   3:32:38.4&  30:58:15.4&  6.4&  2.8& 130.9& 206.& 505.&  5.4&  3\\
 24&   3:32:21.9&  31:22:02.1&  3.3&  2.0&  39.1& 268.& 226.&  6.1&  3\\
 25&   3:30:27.9&  30:26:38.5&  4.5&  1.9& 116.6& 260.& 511.&  4.1&  8\\
 26&   3:29:40.5&  31:37:34.4&  3.9&  2.3&  84.5& 410.& 273.&  4.7&  1\\
 27&   3:29:03.8&  30:04:28.1&  2.7&  2.0&  31.9& 235.& 260.&  5.0&  5\\
 28&   3:28:58.9&  31:22:01.0&  6.5&  3.8& 161.9& 421.& 304.&  7.7&  1\\
 29&   3:28:51.2&  30:44:36.1&  2.0&  2.5&  39.4& 217.& 474.&  1.9& 11\\
 30&   3:28:50.6&  31:09:11.5&  3.3&  3.5&  40.9& 376.& 170.&  3.9&  1\\
 31&   3:28:42.3&  31:12:21.7&  1.4&  3.2&   9.7& 428.&  81.&  0.8&  1\\
 32&   3:28:27.9&  30:19:32.0&  2.3&  2.8&  61.2& 191.& 723.&  1.0&  5\\
 33&   3:27:58.5&  31:26:45.5&  2.8&  2.8&  28.9& 317.& 167.&  4.2&  1\\
 34&   3:27:34.9&  30:11:56.4&  5.1&  3.6&  48.9& 206.& 238.& 10.5&  5\\
 35&   3:27:08.3&  30:05:26.3&  2.4&  2.8&  56.1& 220.& 537.&  1.9&  5\\
 36&   3:26:13.4&  30:29:45.7&  2.6&  2.0&  24.0& 187.& 249.&  5.3& 10\\
 37&   3:25:40.8&  30:09:14.3&  3.0&  1.8&  63.7& 491.& 220.&  2.7&  9\\
 38&   3:25:25.6&  30:42:50.1&  3.5&  2.1&  75.6& 453.& 244.&  3.7& 10\\
 39&   3:24:53.2&  30:22:35.0&  3.2&  2.3&  75.9& 529.& 228.&  2.7&  9\\
\enddata

\tablenotetext{a}{Position of peak extinction within core (accurate to 2.5\arcmin).}
\tablenotetext{b}{Peak extinction, background extinction, mass, $\sigma$'s, and
	mean density derived from results of Gaussian fitting.  See text for details.}
\tablenotetext{c}{Associated extinction super core.}
\end{deluxetable}
\newpage
\begin{deluxetable}{ccccccc}
\tablewidth{0pt}
\tablecolumns{7}
\tablecaption{Properties of extinction super cores in Perseus.
	\label{extsupcore_props}}
\tablehead{
\colhead{Ref \#} &
\colhead{RA \tablenotemark{a}} &
\colhead{Dec \tablenotemark{a}} &
\colhead{Peak \tablenotemark{b}} &
\colhead{Mass \tablenotemark{b}} &
\colhead{R$_{eff}$ \tablenotemark{b}} &
\colhead{$<$n$>$ \tablenotemark{b}}
\\
\colhead{} &
\colhead{(J2000.0)} &
\colhead{(J2000.0)} &
\colhead{A$_{V}$} &
\colhead{(M$_{\odot}$)} &
\colhead{(\arcsec)} &
\colhead{$10^{3}$cm$^{-3}$}
}
\startdata
  1&   3:47:45.3&  32:52:43.4& 10.9&  859.6&  776.&   7.1\\
  2&   3:43:57.1&  31:59:28.7& 10.1& 1938.9& 1119.&   5.3\\
  3&   3:39:27.4&  31:21:08.6& 10.4&  780.6&  737.&   7.5\\
  4&   3:36:28.9&  31:11:13.1&  9.3&  560.5&  670.&   7.2\\
  5&   3:32:35.6&  30:58:27.7&  9.5&  441.1&  579.&   8.8\\
  6&   3:30:28.7&  30:26:30.2&  7.6&  257.6&  454.&  10.6\\
  7&   3:28:56.0&  31:22:36.4&  6.1&  973.3&  889.&   5.3\\
  8&   3:28:53.3&  30:44:00.5&  7.0&  246.2&  453.&  10.2\\
  9&   3:27:36.6&  30:12:32.8&  6.1&  240.1&  448.&  10.3\\
 10&   3:25:22.8&  30:43:19.2&  5.9&  173.7&  386.&  11.6\\
 11&   3:24:50.3&  30:23:10.1&  5.6&  107.4&  309.&  14.0\\
\enddata

\tablenotetext{a}{Position of peak extinction within core (accurate to 2.5\arcmin).}
\tablenotetext{b}{Peak extinction, mass, radius, and mean number density 
	derived from Clumpfind 
	\citep{clfind} with several clumps further separated.  See text for details. }

\end{deluxetable}


\begin{deluxetable}{ccccccccc}
\tabletypesize{\small}
\tablewidth{0pt}
\tablecolumns{9}
\tablecaption{Distribution of mass in the Perseus molecular cloud binned with extinction.\label{ext_bin}}
\tablehead{
\colhead{A$_{V}$} & 
\colhead{Cloud Area \tablenotemark{a}} & 
\multicolumn{2}{c}{Cloud Mass \tablenotemark{a}} & 
\multicolumn{2}{c}{Cloud Mass \tablenotemark{b}} & 
\multicolumn{2}{c}{Clump Mass} & 
\colhead{Mass Ratio}\tablenotemark{b} \\

\colhead{Range} &
\colhead{($\%$)} &
\colhead{M$_{\odot}$} &
\colhead{$\%$} &
\colhead{M$_{\odot}$} &
\colhead{$\%$} &
\colhead{M$_{\odot}$} &
\colhead{$\%$} &
\colhead{($\%$)}  
}

\startdata

0-12 & 100  & 18572 & 100  & 6094  & 100  & 61.3 & 100 & 1.0 \\
0-5  & 95.3 & 15842 & 85.3 & 3515  & 57.7 & 0.6  & 1.0 & 0   \\
5-10 &  4.6 & 2652  & 14.3 & 2502  & 41.1 & 53.0 & 86.4& 2.1 \\
10-12&  0.1 & 78    & 0.4  & 78    & 1.2  & 7.7  & 12.6& 9.9 \\
\enddata

\tablenotetext{a}{Over entire area of our extinction map}
\tablenotetext{b}{Over the region of the extinction map where
	submillimetre data also exist}

\end{deluxetable}

\begin{figure}[p]
\centerline{\includegraphics[width=16cm]{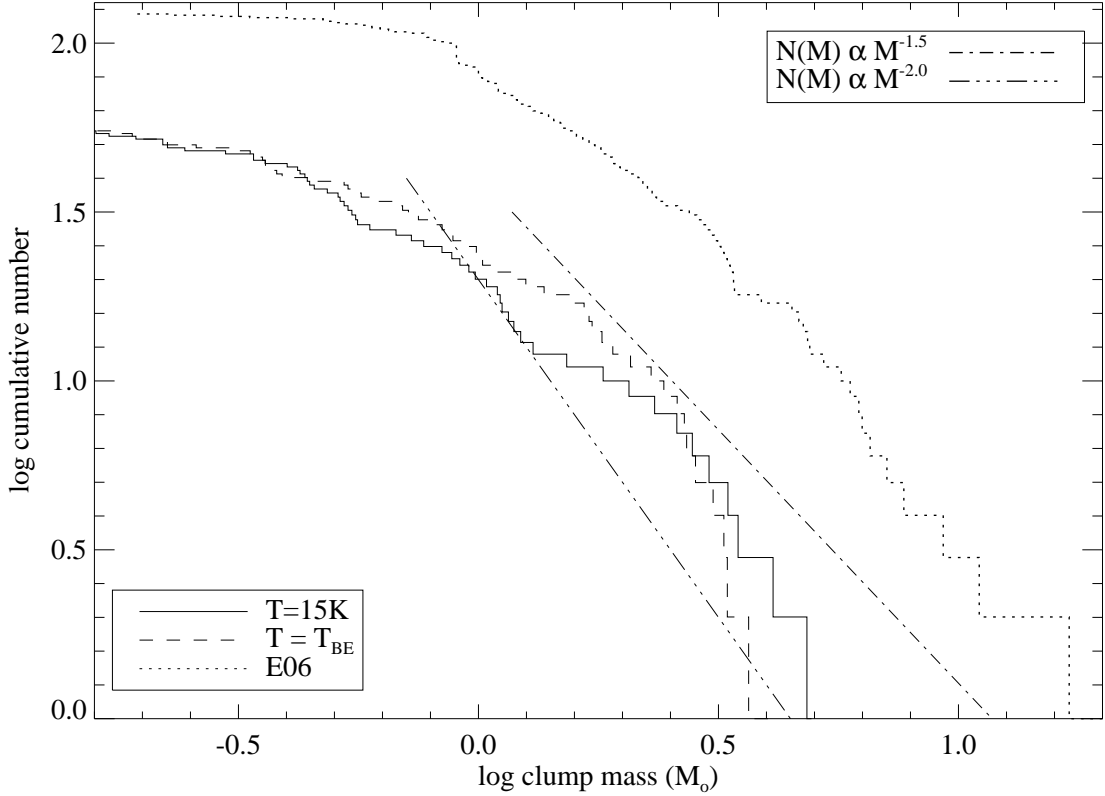}}
  \caption{Cumulative mass distribution of the submillimetre clumps showing
        mass estimates using both a constant (T = 15~K) and Bonnor-Ebert
        model fit temperature.  Also plotted (E06) is the mass distribution found
        by \citet{Enoch06} in a 1.1~mm survey of the Perseus cloud, which
        had both a larger areal coverage and sensitivity to larger objects,
        leading to the offsets in the mass distributions.
	Mass distribution slopes of -1.5 and -2 are shown to guide 
	the eye.}
  \label{massdistrib_submm}
\end{figure}

\end{document}